\documentclass[aps,twocolumn,nofootinbib,showpacs,prd,10pt,superscriptaddress]{revtex4-2}
\usepackage{graphicx}
\usepackage[english]{babel}
\usepackage[T1]{fontenc}
\usepackage{mathrsfs}
\usepackage[tbtags]{amsmath}
\usepackage{amssymb}
\usepackage{amsxtra}
\usepackage{amsopn}
\usepackage{latexsym}
\usepackage[mathcal]{eucal}
\usepackage{mathtools}
\usepackage{slashed}
\usepackage{xcolor}
\usepackage[hyperfootnotes=false,bookmarks=false]{hyperref}

\usepackage{braket}
\usepackage{enumerate}

\newcommand{\BE}{\begin{equation}}
\newcommand{\EE}{\end{equation}}
\newcommand{\BA}{\begin{align}}
\newcommand{\EA}{\end{align}}

\newcommand{\mc}{\mathcal}
\newcommand{\psibar}{\overline{\psi}}
\newcommand{\cbar}{\overline{c}}

\begin{document}

\title{One-loop analytic structure of the deep-infrared Landau-gauge\\gluon propagator at finite temperature}

\author{Giorgio Comitini}
\email{giorgio.comitini@dfa.unict.it}
\affiliation{Dipartimento di Fisica e Astronomia ``E. Majorana'', Universit\`a di Catania, Via S. Sofia 64, I-95123 Catania, Italy}
\affiliation{INFN Sezione di Catania, Via S. Sofia 64, I-95123 Catania, Italy}

\date{\today}

\begin{abstract}
    With the aim of looking for signatures of deconfinement, the poles and the spectral function of the Landau-gauge gluon propagator are investigated at one loop, vanishing spatial momentum, and finite temperature within the framework of the screened massive expansion of pure Yang-Mills theory and of full QCD. When computed using both temperature-independent parameters optimized by principles of gauge invariance and temperature-dependent parameters obtained by fitting lattice data at zero Matsubara frequency, the propagator is found to have a pair of complex-conjugate poles in its squared complexified frequency variable throughout the considered temperature interval, ranging from $T=0$ to temperatures $T>T_{c}$ of interest to quark-gluon plasma phenomenology. The spectral function is found to violate positivity and not to develop sharp peaks over said temperature interval. In full QCD, a simple model is used for mass generation in the infrared quark sector; the dependence of our results on the quark masses is discussed.
\end{abstract}

\maketitle
\section{Introduction}

If gluons were physical degrees of freedom living in a positive-definite Hilbert space, their propagator would admit an ordinary K\"{a}ll\'{e}n-Lehmann representation, allowing it to be expressed as an integral over a positive-definite spectral function. Gluon asymptotic states would then show up as Dirac deltas in the spectral function, whereas finite-lifetime excitations would manifest themselves as sharp but finite spectral peaks. In both cases, physical excitations of the gluon field would correspond to poles in the gluon propagator: the deltas would originate from poles on the Minkowski axis of the propagator, whereas the finite peaks would be the image of complex poles in secondary Riemann sheets of the propagator \cite{BIR06} -- reached by analytically continuing the propagator beyond the logarithmic branch cut on its Minkowski axis -- just outside of its physical (principal) Riemann sheet\footnote{Counterexamples to this one-to-one correspondence are however known to exist, see e.g. \cite{BI97}.}. Within this picture, understanding the physical nature of the gluonic degrees of freedom would be as easy~-- or, should we rather say, as difficult -- as computing the singularities or the spectral function of the exact gluon propagator. For instance, if the propagator was found not to have poles nor peaks in its spectral function at sufficiently low temperatures, that could be interpreted as direct proof of confinement.

Unfortunately, in the case of non-abelian gauge theories such as QCD, nature has not been so kind as to allow for such a simple characterization of the gauge degrees of freedom. On the one hand, the requirement of explicit Lorentz covariance forces representations of the gluon fields' commutator algebra to live in a Hilbert space with indefinite metric \cite{Gup50,Bleu50,Nak72,KO79} which contains states of zero and negative norm. In principle, not only can these states yield negative contributions to the gluon spectral function, but they are also permitted to have complex energy eigenvalues -- coming in conjugate pairs to preserve the hermiticity of the Hamiltonian --, which in turn would manifest themselves as complex-conjugate poles in the \textit{physical} Riemann sheet of the propagator \cite{HK21b,SC22a}; both of these features are incompatible with the ordinary K\"{a}ll\'{e}n-Lehmann spectral representation. On the other hand, standard calculations carried out in ordinary perturbation theory tell us that -- owing to the asymptotic freedom of non-abelian gauge theories -- the one-loop anomalous dimension of the gluon propagator is negative. This in turn implies that, in the UV, the gluon spectral function \textit{is} negative \cite{OZ80}, and thus violates the constraints of the ordinary K\"{a}ll\'{e}n-Lehmann representation not only in principle, but also in practice. A multitude of non-perturbative approaches, encompassing both lattice simulations \cite{CMT05,SIMS06,SO06,BHLP07,CM10} and continuum studies \cite{ADFM04,HAH13,SIR16b,SC18,KWHM20}, have by now confirmed that positivity violation also affects the infrared regime of the theory.

Complex-conjugate massive poles in the physical sheet of the vacuum gluon propagator made their first appearance in the context of the Gribov-Zwanziger scenario \cite{Zwa89} and, later on, of its refined extension accounting for gauge condensates \cite{DGSVV08}. They are routinely used to describe the deep-infrared behavior of the gluon propagator simulated on the lattice \cite{CM10,DOV10,CDMV12}, they show up in some reconstructions of the gluon spectral function from Euclidean data \cite{BT20} and they can be explicitly obtained by straightforward one-loop calculations within perturbative massive models of QCD \cite{SIR16b,SIR17d,SC18,HK19,HK20,SC21,HK21a}. Despite their ubiquitousness in studies of the infrared regime of QCD, our understanding of their nature remains seriously hindered by the current lack of a complete and consistent field-theoretical framework in which they coexist with non-negotiable requirements of Lorentz covariance, unitarity, causality and the existence of a momentum-space representation of the theory in Minkowski space \cite{Nak71a,Nak71b,Sti94,HK21b,ABCM25}. The fact that they are not associated to physical particles in any ordinary sense -- if this was the case, as we said, they would be located either on the Minkowski axis, or closeby, in secondary Riemann sheets of the propagator~-- is usually interpreted as a symptom of gluon confinement \cite{Sti94,SIR17a,HK19,BT20}. The fact that they are exactly gauge invariant over a wide class of gauges \cite{Niel75b,KKR90,BLS95} signals that their position in the complex plane may directly enter the observables of the theory, and makes them worthy of in-depth investigations.

Over the past fifteen years, the realization that gluons acquire a finite mass in the infrared \cite{LSWP98a,LSWP98b,BBLW00,BBLW01,AN04,BHLP04,SO04,SIMS05,SO06,AP06,BHLP07,IMSS07,ABP08,AP08,BIMS09,ISI09,BMM10,SO10,OS12,ABBC12,DOS16}, which in turn protects the QCD coupling constant from developing a Landau pole and keeps it moderately small at all energy scales \cite{SIMS05,AP06,IMSS07,AP08,BIMS09,DOS16}, has led to the development of multiple independent perturbative massive models for the infrared behavior of QCD \cite{TW10,TW11,SIR16a,CFGM16b,CvEPG18}. These models have proven good beyond expectations in reproducing the low-energy lattice data for the Green functions of the theory \cite{TW10,TW11,PTW13,PTW14,PTW15,SIR16a,CDDS24}, and have also been employed to study QCD at finite temperatures and densities \cite{RSTW14,RST15,RSTW15a,RSTW16,SIR17d,RSTT17,CS18,HK21a,PRST21a,SC21,CS25}. Among all approaches to infrared QCD, they occupy a unique position in that -- especially to low loop order -- they are able to provide analytic expressions which can be easily continued to the complex energy plane, giving access to both Euclidean and Minkowski space and enabling the direct study of the singularities of the Green functions of the theory.

Of particular interest in this context is the topic of the evolution of the gluon poles with temperature. Both pure Yang-Mills theory and full QCD are indeed known to undergo a phase transition\footnote{Which in the case of full QCD with sufficiently light quarks is actually a smooth crossover.} beyond which gluons and quarks are expected to behave as deconfined -- and, at sufficiently high temperatures, weakly interacting -- quasi-particles, making up for the state of matter known as the quark-gluon plasma. The natural question then arises of what happens to the gluon poles as temperature increases beyond the deconfinement temperature $T_{c}$: by virtue of the information propagators usually hold on quasi-particles, the transition from confined to deconfined gluons can be reasonably expected to reflect itself into a change in the analytic structure of the gluon propagator.

An attempt to address this question \cite{SIR17d,SC21} was made in the Landau gauge within the framework of the so-called screened massive expansion \cite{SIR16a,SIR16b,SIR17a,SIR17b,SIR17c,SIR17d,CS18,SC18,SIR19a,SIR19b,CS20,CRBS21,SC21,SC22a,SC22b,SIR23,SC23,CS25}, a perturbative approach to infrared QCD based on the idea that a suitable choice of the zero order of perturbation theory reflecting dynamical mass generation in the transverse gluon sector may include gluon mass effects into QCD's perturbative series and minimize the influence of higher-loop corrections without changing QCD's standard Faddeev-Popov action. Within the screened massive expansion, at one loop, it was found that, for sufficiently high temperatures $T$, the real and the imaginary part of the complex gluon poles start to increase linearly with $T$ \cite{SIR17d,SC21}, a behavior which is fully expected from quasi-particle poles\footnote{Modulo logarithmic corrections due to the running of the coupling constant with temperature.}. One aspect of the calculation that was not sufficiently emphasised, however, is that said poles, at face value, are \textit{not} those one would usually associate to ordinary physical quasi-particles in the deconfined phase. These thermal poles, in fact, evolve straight from those one computes to one loop in zero-temperature massive models of QCD: they are still complex conjugate, and they are still located in the principal Riemann sheet of the gluon propagator, meaning that their naive interpretation should be closer to that we customarily give to massive complex-conjugate poles in vacuum.

While it cannot be excluded \textit{a priori} that the thermal poles computed in the screened massive expansion at one loop may be related to physical gluon quasi-particles~-- we will have the chance to address this topic in more depth in the conclusions --, the absence of a clear signature of deconfinement in terms of changes in the pole structure of the gluon propagator fuels the need to look for evidence of deconfinement also in other directions. In this respect, a natural next step would be to, quite literally, look for such evidence in the \textit{opposite} direction to complex-conjugate poles: instead of outward in the complex plane, one could turn their attention back towards the Minkowski axis of the propagator, and compute the gluon spectral function at finite temperature. A clear signature of deconfinement there would be for instance the formation of a sharp quasi-particle peak at temperatures greater than $T_{c}$. Such a peak may enter physical observables, e.g., via the retarded gluon propagator in linear response theory and could be directly computed within perturbative massive models, assuming that the approximation is sufficiently accurate.

To our knowledge, there are only a limited number of works discussing the gluon's spectral function at finite temperature in the literature. Those we are aware of \cite{HFP14,SDO14,SODBC14,CHPS15,SODR17a,SODR17b,IPRT18} all involve solving the ill-defined problem of computing the spectral function from a finite set of data for the gluon propagator in Euclidean space, which is in turn obtained either by functional Renormalization Group methods \cite{HFP14,CHPS15} or from lattice simulations \cite{SDO14,SODBC14,SODR17a,SODR17b,IPRT18}. By converse, we are not aware of any calculation carried out directly in the complex plane within massive perturbative models of QCD\footnote{For results obtained at finite quark density in the Curci-Ferrari model see however \cite{HK21a}.}. The main objective of this paper is to bridge this gap and present results on the finite-temperature gluon spectral function computed in the deep infrared within the framework of the screened massive expansion of both pure Yang-Mills theory and of full QCD, at one loop and in the Landau gauge. While the calculation could be easily extended to finite spatial momenta $|{\bf p}|$, in the present paper we will limit ourself to displaying results at vanishing momentum $|{\bf p}|=0$. Next to the spectral function, we will also display results on the complex-conjugate gluon poles as a function of temperature. Similar results, as we mentioned, were already reported for pure Yang-Mills theory in \cite{SIR17d,SC21}\footnote{As already noted in \cite{CS25}, Ref.~\cite{SC21} contains an error in the parameters used to fit the pure Yang-Mills Euclidean propagators of \cite{SOBC14} in the static limit. The error was corrected in \cite{CS25} for what pertains the propagators themselves, and in the present paper we report the corresponding corrected pure Yang-Mills theory poles.}, whereas those computed in full QCD are presented here for the first time.

This paper is organized as follows. In Sec.~II we will describe the setup of the screened massive expansion in pure Yang-Mills theory and full QCD, in vacuum and at finite temperature. In Sec.~III and in Sec.~IV we will display the zero-momentum poles and spectral function of the Landau-gauge gluon propagator at finite temperature, respectively, in pure Yang-Mills theory and in full QCD. Finally, in Sec.~V, we discuss our results and present our conclusions.

\section{The screened massive expansion at zero and finite temperature}

\subsection{Setup}

The screened massive expansion \cite{SIR16a,SIR16b,SIR17a,SIR17b,SIR17c,SIR17d,CS18,SC18,SIR19a,SIR19b,CS20,CRBS21,SC21,SC22a,SC22b,SIR23,SC23,CS25} is a deformation of ordinary QCD perturbation theory formulated with the aim of treating the transverse gluons as massive already at tree level. Denoting the renormalized Euclidean pure Yang-Mills Faddeev-Popov Lagrangian in a generic linear covariant gauge with $\mc{L}_{\text{YM}}$,
\begin{align}\label{fplag}
    \mathcal{L}_{\text{YM}} & =\frac{1}{4}\,F_{\mu\nu}^{a}F^{a\,\mu\nu}+\frac{1}{2\xi}\,(\partial\cdot A^{a})^{2}+\cbar^{a}\partial\cdot D c^{a}=                                       \\
    \notag                  & =\frac{1}{2}\,\partial_{\mu}A_{\nu}^{a}\left[\partial^{\mu}A^{a\,\nu}-\left(1-\frac{1}{\xi}\right)\partial^{\nu}A^{a\,\mu}\right]+\overline{c}^{a}\partial^{2}c^{a}+ \\
    \notag                  & \quad+gf^{a}_{bc}\,\partial^{\mu}A^{a\,\nu}A_{\mu}^{b}A_{\nu}^{c}+\frac{g^{2}}{4}f^{a}_{bc}f^{a}_{de}A_{\mu}^{b}A_{\nu}^{c}A^{d\,\mu}A^{e\,\nu}+          \\
    \notag                  & \quad+gf^{a}_{bc}\,\partial^{\mu}\overline{c}^{a}c^{b}A_{\mu}^{c}\ ,
\end{align}
where $F_{\mu\nu}^{a}=\partial_{\mu}A_{\nu}^{a}-\partial_{\nu}A_{\mu}^{a}+gf^{a}_{bc}A_{\mu}^{b}A_{\nu}^{c}$ is the gluon field-strength tensor, $D_{\mu}^{ab}=\delta^{ab}\partial_{\mu}-f^{abc}A_{\mu}^{c}$ is the SU($N$) covariant derivative acting on the adjoint representation, $f_{abc}$ are the SU($N$) structure constants, $g$ is the strong coupling, $\xi$ is the gauge parameter selecting the linear covariant gauge, and we have omitted the standard renormalization counterterms, the screened massive expansion is defined by choosing as the order zero of perturbation theory the quadratic action $S_{\text{YM},m}$,
\begin{equation}\label{Sm}
    S_{\text{YM},m}=\int d^{d}x\ \left\{\mc{L}_{\text{YM},0}+\frac{1}{2}\,m^{2}\,A_{\mu}^{a}\,t^{\mu\nu}A_{\nu}^{a}\right\}\ ,
\end{equation}
where $\mc{L}_{\text{YM},0}$ is the ordinary (massless) kinetic Lagrangian,
\begin{equation}
    \mc{L}_{\text{YM},0}=\frac{1}{2}\,\partial_{\mu}A_{\nu}^{a}\left[\partial^{\mu}A^{a\,\nu}-\left(1-\frac{1}{\xi}\right)\partial^{\nu}A^{a\,\mu}\right]+\overline{c}^{a}\partial^{2}c^{a}\ ,
\end{equation}
$m^{2}$ is a (squared) mass parameter and $t_{\mu\nu}$ is the transverse projector -- in momentum space,
\begin{equation}
    t_{\mu\nu}(p)=\delta_{\mu\nu}-\frac{p_{\mu}p_{\nu}}{p^{2}}\ .
\end{equation}
The mass term in Eq.~\eqref{Sm} causes the transverse component of the bare Euclidean gluon propagator $\Delta_{m,ab}^{\mu\nu}(p)$ derived from $S_{\text{YM},m}$,
\begin{equation}\label{glupropzero}
    \Delta_{m,ab}^{\mu\nu}(p)=\delta_{ab}\left(\frac{t^{\mu\nu}(p)}{p^{2}+m^{2}}+\xi\,\frac{\ell^{\mu\nu}(p)}{p^{2}}\right)\ ,
\end{equation}
where $\ell_{\mu\nu}(p)=p_{\mu}p_{\nu}/p^{2}$ is the longitudinal projector, to acquire a mass $m$, while leaving both its longitudinal component and the bare ghost propagator $\mc{G}_{0}(p)$,
\begin{equation}
    \mc{G}_{0}(p)=\frac{1}{-p^{2}}\ ,
\end{equation}
massless. Since the mass term was not present in the original Faddeev-Popov action, in order for the latter to remain unaffected by its introduction, the same term is subtracted back from the interaction Lagrangian $\mc{L}_{\text{int},m}$:
\begin{equation}
    \mc{L}_{\text{int},m}=\mc{L}_{\text{int},0}-\frac{1}{2}\,m^{2}\,A_{\mu}^{a}\,t^{\mu\nu}A_{\nu}^{a}\ ,
\end{equation}
where $\mc{L}_{\text{int},0}$, the ordinary Faddeev-Popov interaction Lagrangian, contains the three-gluon, four-gluon and ghost-gluon interaction vertices, plus the standard renormalization counterterms. The new term in $\mc{L}_{\text{int},m}$ yields a two-gluon interaction vertex $\delta\Gamma^{ab}_{\mu\nu}(p)$ -- termed the gluon ``mass counterterm''\footnote{Not to be confused with a renormalization counterterm.} --,
\begin{equation}\label{gmct}
   \delta\Gamma^{ab}_{\mu\nu}(p)=m^{2}\,t_{\mu\nu}(p)\,\delta^{ab}\ ,
\end{equation}
which must be included in the Feynman rules of the screened massive expansion.

In order to compute quantities of interest within the framework of the screened massive expansion, one uses the standard perturbative techniques. Nevertheless, the method is non-perturbative in nature, in that the mass parameter $m^{2}$ that appears in $\delta\Gamma_{\mu\nu}^{ab}(p)$ does not originate from ordinary perturbative calculations and is not taken to be proportional to a fixed power of the coupling constant $g$. It should be noted that the introduction of $m^{2}$ in the action of the theory reduces its predictivity by making the resulting perturbative series depend on an extra free parameter in comparison to ordinary perturbation theory. The screened massive expansion must thus be supplemented both with a summation scheme for the so-called ``crossed diagrams'' -- that is, diagrams containing one or more crossed counterterms -- and with a way to reduce the number of free parameters of the expansion. These issues will be addressed in the next section when we will review the calculation of the gluon propagator.

The screened massive expansion can be applied both to pure Yang-Mills theory and to full QCD. When working with the latter, one finds \cite{SIR16b,CRBS21} that the corrections to the perturbative series brought by the introduction of the gluon mass do not generate an infrared mass for chiral quarks, nor do they considerably enhance the mass of light quarks\footnote{At least in the context of one-loop fixed-scale calculations. The renormalization group flow of full QCD -- and in particular the running of the quark mass parameter -- was never studied within the framework of the screened massive expansion, nor were higher-loop calculations ever performed.}. It follows that the phenomenon of chiral symmetry breaking and the subsequent generation of a mass of the order of the QCD scale $\Lambda_{\text{QCD}}$ which is known to characterize the infrared dynamics of the quark sector \cite{KBLW05} must be addressed by separate methods.

One method that proved successful in describing the light quarks' momentum-dependent mass function at low energies \cite{CRBS21} consists in performing, in the quark sector, a shift of the perturbative expansion point analogous to the one carried out in the gluon sector: denoting the Euclidean quark Lagrangian with $\mc{L}_{q}$,
\begin{equation}
    \mc{L}_{q}=\sum_{f}\psibar_{f}(\slashed{D}+m_{f})\psi_{f}\ ,
\end{equation}
where the sum is over quark flavors $f$, $D_{\mu}=\partial_{\mu}-igA_{\mu}^{a}T_{a}$~-- $T_{a}$ being SU($N$) generators~-- is the covariant derivative in the fundamental representation and the $m_{f}$'s are current quark masses, one can choose as the zero-order fermionic action $\mc{L}_{q,M}$
\begin{equation}
    \mc{L}_{q,M}=\sum_{f}\psibar_{f}(\slashed{\partial}+M_{f})\psi_{f}\ ,
\end{equation}
where the $M_{f}$'s are the quark masses generated in the infrared by chiral symmetry breaking; modulo renormalization counterterms, the quark interaction Lagrangian $\mc{L}_{q,\text{int}}$ will then read
\begin{equation}\label{qmint}
    \mc{L}_{q,\text{int}}=\sum_{f}\psibar_{f}[-ig\slashed{A}-(M_{f}-m_{f})]\psi_{f}\ .
\end{equation}
$\mc{L}_{q,\text{int}}$ contains new non-perturbative two-point quark vertices, proportional to the mass differences $M_{f}-m_{f}$, which are entirely analogous to the gluon mass counterterm $\delta\Gamma_{\mu\nu}^{ab}(p)$ in Eq.~\eqref{gmct}.

Since the objective of this paper is to explore the analytic structure of the gluon propagator, rather than the quark sector, in what follows we will employ a simpler, less sophisticated model for infrared quarks. Instead of keeping the full quark interaction as in Eq.~\eqref{qmint}, we truncate it to only contain the quark-gluon vertices. Equivalently, we treat quarks as if the masses with which they propagate are the infrared $M_{f}$'s, with no further corrections. In practice, we take their zero-order propagator $S_{f,M}(p)$ to be
\begin{equation}
    S_{f,M}(p)=\frac{1}{i\slashed{p}+M_{f}}\ ,
\end{equation}
where $M_{f}\sim\Lambda_{\text{QCD}}$, and keep their interactions unchanged from those of ordinary perturbative QCD.

This model was already employed in \cite{CS25} to study the behavior of the gluon propagator at finite temperature and density in full QCD. There we discussed the validity of the approximation in light of restoration of chiral symmetry, known to take place in the deconfined phase of QCD for temperatures $T>T_{c}\approx 150$~MeV \cite{Baz12}. By virtue of chiral symmetry restoration, the quark mass scale is expected to decrease beyond the deconfinement crossover. At larger temperatures, it is then expected to increase with $T$ due to thermal effects, as predicted by ordinary thermal perturbative QCD. To date, a quantitative description of the temperature dependence of the quark mass scale is still unavailable within the framework of the screened massive expansion. As a consequence, we are unable to specialize our model so that chiral symmetry restoration and thermal effects are properly accounted for. To mitigate this deficiency in the context of the present study, in a later section we will explore the effects produced by a decrease in the quark masses on the poles and spectral function of the gluon propagator.\\

The screened massive expansion can be extended to finite temperatures \cite{SIR17d,CS18,SC21,CS25} by making use of the Matsubara formalism \cite{KG23}. The latter is obtained by restricting the domain of integration of the Euclidean action to the imaginary time interval $\tau\in [0,\beta]$ -- with $\beta=1/T$ the inverse temperature of the system~-- and by imposing periodic (resp. antiperiodic) boundary conditions in imaginary time for the gluon and ghost (resp. quark) fields. In momentum space, this is equivalent to replacing integrals over the fourth component $q^{4}$ of momentum~-- which now can take up values $q^{4}=\omega_{n}=2\pi n T$ for gluons and ghosts and $q^{4}=\omega_{n}=(2n+1)\pi T$ for quarks, $n\in\mathbb{Z}$ -- with discrete sums,
\begin{equation}
    \int\frac{d q^{4}}{2\pi}\to T\sum_{n}\ .
\end{equation}
These sums -- termed Matsubara sums -- can be transformed back to integrals involving the Bose and Fermi distributions \cite{KG23} using the identities
\begin{align}\label{msumb}
    T\sum_{n}f(\omega_{n})&=\int\frac{dp^{4}}{2\pi}\frac{f(p^{4})+f(-p^{4})}{2}+\\
    \notag&\quad+\int_{-\infty+i\epsilon}^{+\infty+i\epsilon}\frac{dz}{2\pi}\frac{f(z)+f(-z)}{e^{-i\beta z}-1}
\end{align}
if $\omega_{n}=2\pi n T$ (bosonic sums), or
\begin{align}\label{msumf}
    &T\sum_{n}f(\omega_{n})=\int_{-\infty}^{+\infty}\frac{dp^{4}}{2\pi}\ f(p^{4})+\\
    \notag&\qquad\qquad\qquad-\int_{-\infty+i\epsilon}^{+\infty+i\epsilon}\frac{dz}{2\pi}\frac{f(z)+f(-z)}{e^{-i\beta z}+1}\ ,
\end{align}
if $\omega_{n}=(2n+1)\pi T$ (fermionic sums), where $f(z)$ on the right-hand side of the above equations is the (unique) analytic continuation of $f(\omega_{n})$ to the complex plane, $z\in\mathbb{C}$. Because of the different footing on which the (imaginary) temporal and spatial directions are treated in the Matsubara formalism, Euclidean O(4) invariance is broken at its most fundamental level to O(3) in the thermal theory. Ultimately, this effect stems from the existence of a preferred reference frame at finite temperature -- namely, that of the thermal medium at rest.

The screened massive expansion as presented in this paper was formulated with the aim of providing an optimized perturbation theory tailored to the deep-infrared behavior of gluons in the vacuum. The breakage of O(4) invariance at finite temperature adds one new degree of complexity to the picture: at $T\neq 0$, the chromomagnetic (i.e., spatially transverse) gluon degrees of freedom behave differently than the chromoelectric (i.e., spatially longitudinal) ones. This property is not specific to the gauge sector of QCD, but applies to all vector bosons at finite temperature, already at the perturbative level \cite{KG23}. In the context of QCD, this difference becomes especially marked in the static limit ($p^{4}=\omega_{n}=0$), where Landau-gauge lattice calculations carried out in pure Yang-Mills theory \cite{SOBC14} reveal that, while the chromomagnetic mass essentially increases with the temperature $T$ -- displaying a behavior consistent e.g. with the predictions of the Gribov-Zwanziger scenario \cite{Zwa06} --, the chromoelectric one experiences a large decrease as $T$ approaches the deconfinement temperature $T_{c}$, to then start increasing only when $T>T_{c}$.

Whether the screened massive expansion in the present form is able to account for such a marked difference was investigated first in \cite{SC21} and, more recently, in \cite{CS25}, where the calculations were also extended to full QCD and to finite baryonic densities. There we found that, in the static limit of the Landau gauge, the formalism does indeed reproduce the qualitative behavior described above. However, quantitatively, the drop in the chromoelectric mass predicted by the screened massive expansion is nowhere nearly as large as that observed for pure Yang-Mills theory on the lattice. Additionally, in order to reproduce the chromoelectric sector around the critical temperature, one is forced to use a different set of parameters than in the chromomagnetic sector, and even then predictions fail for momenta lower than $\approx0.5$-$0.7$ GeV. As for full QCD, preliminary results \cite{Com25b} comparing the predictions of the screened expansion to recently reported unquenched $n_{f}=2$ lattice data \cite{SOS25} suggest that, at finite temperature, just as in pure Yang-Mills theory, the method works better in the chromomagnetic sector than in the chromoelectric one, at least to one loop. Nonetheless, a milder decrease of the chromoelectric mass towards $T_{c}$ with respect to the pure Yang-Mills case~-- presumably due to the sharp deconfinement transition becoming a crossover in full QCD~-- contributes to considerably improving the agreement of the screened massive expansion with the lattice data also in the static chromoelectric sector. It does not, however, eliminate the need to use distinct parameters from the chromomagnetic sector.

These shortcomings of the screened massive expansion suggest that, at finite temperature, shifting the zero-order action of full QCD by a (four-dimensionally) transverse gluon mass term as one does in Eq.~\eqref{Sm} may constitute a suboptimal ansatz\footnote{Whether or not the agreement with the lattice data would improve by going to higher order in perturbation theory is currently unknown.}. A more accurate description of the infrared behavior of gluons at $T\neq 0$ may be provided by a gluon mass counterterm that accounts for the different behavior of the chromoelectric and the chromomagnetic masses -- as it happens, for instance, implicitly in the framework of the hard-thermal loop (HTL) screened perturbation theory \cite{ABS00}. Still, there are a number of reasons why it makes sense to continue using the simple ansatz provided by Eq.~\eqref{Sm}. First of all, treating the chromoelectric and the chromomagnetic mass differently requires a first-principles description of such a difference, which is not currently available, if not only qualitatively. Second of all, lattice results at non-zero Matsubara frequencies \cite{SODR18} suggest that, outside of the static limit, the asymmetry between the chromoelectric and the chromomagnetic sectors is not as marked as in said limit, so that calculations carried out at $p^{4}\neq0$ can be expected to be less affected by it. Third, the chromomagnetic sector in and of itself is described sufficiently well by Eq.~\eqref{Sm}, even quantitatively, at $p^{4}=\omega_{n}=0$, when temperature-dependent parameters (see the next section) are used. Finally, in the present study, we are interested in the deep-infrared behavior of the gluon propagator at non-zero, complex frequencies -- $p^{4}\in\mathbb{C}$, $|{\bf p}|\to 0$. It can be shown analytically \cite{SC21,KG23} that, for $p^{4}\neq 0$, the chromoelectric and the chromomagnetic components of the gluon propagator reduce to the same function when $|{\bf p}|=0$. In this regime, therefore, the asymmetry between the two sectors can clearly play no role. Indeed, the results reported in Secs.~III and IV will only feature a single set of poles and a single spectral function for both the chromoelectric and the chromomagnetic sector.

\subsection{Gluon propagator}

The vacuum gluon propagator was first computed using the screened massive expansion of pure Yang-Mills theory in \cite{SIR16a} and of full QCD in \cite{SIR16b}; it was later computed at finite temperature in pure Yang-Mills theory in \cite{SC21}, and, recently, at finite temperature and baryonic density in full QCD in \cite{CS25}. As noted in the previous section, due to its non-perturbative nature, the expansion must be supplemented by a summation scheme for the crossed diagrams~-- that is, of diagrams containing one or more insertions of the gluon mass counterterm. In more detail, diagrams which only differ by the number of mass counterterms in their internal gluon lines are of the same order in the coupling $g$, which means that, in principle, they could all be resummed at any fixed order in perturbation theory. Such a resummation can be easily shown \cite{SIR16a} to lead back to ordinary (massless) perturbation theory, and must thus be avoided.

\begin{figure}[h]
    \centering
    \includegraphics[width=0.10\textwidth]{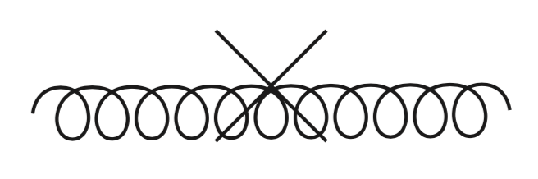}
    \vskip 5pt
    \includegraphics[width=0.45\textwidth]{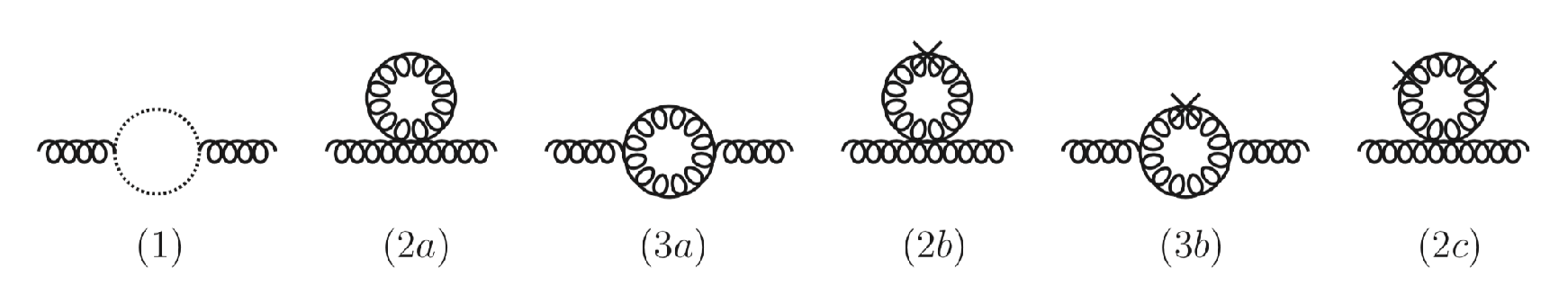}
    \caption{1PI diagrams with no more than three vertices used to compute the one-loop gluon polarization in the screened massive expansion of pure Yang-Mills theory. Top: tree-level gluon mass counterterm. Bottom: loop diagrams. A diagram specular to (3b) with one mass counterterm in the lower internal gluon line is not displayed explicitly, but included in (3b) itself.}
    \label{glu1ldiags}
\end{figure}

In order to determine the number of crossed diagrams to retain at any fixed order in perturbation theory, one can resort to principles of minimality and renormalizability. Let us first consider the ordinary one-loop diagrams making up the gluon propagator in perturbative Yang-Mills theory, diagrams (1), (2a) and (3a) in Fig.~\ref{glu1ldiags}. The ghost loop -- diagram (1) -- contains no internal gluon lines and thus evaluates to its counterpart in ordinary perturbation theory. On the other hand, because of the extra mass term in Eq.~\eqref{Sm}, in the screened massive expansion diagrams (2a) and (3a) must be computed with a massive bare transverse gluon propagator. As a consequence, diagram (2a) doesn't vanish at zero temperature as it happens in dimensionally regularized ordinary perturbation theory. On the contrary, together with diagram (3a), it contains a mass divergence -- that is, a divergence proportional to $m^{2}$ -- which cannot be eliminated by the renormalization counterterms of ordinary perturbation theory. Diagrams (2b) and (3b), however, are easily seen to contain the same mass divergences as (2a) and (3a), respectively, with an opposite sign \cite{SIR16a,SC18}: since inserting a single gluon mass counterterm is equivalent to making the replacement
\begin{equation}
    \frac{1}{p^{2}+m^{2}}\to \frac{m^{2}}{(p^{2}+m^{2})^{2}}=-m^{2}\frac{\partial}{\partial m^{2}}\left(\frac{1}{p^{2}+m^{2}}\right)
\end{equation}
in the transverse component of the bare propagator, denoting the contribution of diagram $(i)$ to the gluon polarization with $\Pi_{\mu\nu}^{(i)}(p)$, one has
\begin{equation}
    \Pi_{\mu\nu}^{(2b,3b)}(p)=-m^{2}\frac{\partial}{\partial m^{2}}\,\Pi_{\mu\nu}^{(2a,3a)}(p)\ .
\end{equation}
Clearly, any term linear in $m^{2}$ in $\Pi_{\mu\nu}^{(2a,3a)}(p)$ -- including mass divergences -- appears in $\Pi_{\mu\nu}^{(2b,3b)}(p)$ with the opposite sign. Given that the derivative kills the ordinary $p^{2}$ divergence in $\Pi_{\mu\nu}^{(2a,3a)}(p)$, the sums $\Pi_{\mu\nu}^{(2a,3a)}(p)+\Pi_{\mu\nu}^{(2b,3b)}(p)$ only contain the ordinary gluon field-strength divergence, proportional to $p^{2}$, which can be eliminated by the usual gluon field-strength renormalization counterterm. Including diagrams (2b) and (3b) in the one-loop polarization is thus necessary to ensure the renormalizability of the propagator.

Once diagram (3b) is added to the one-loop polarization, for consistency, we also include all and only the remaining diagrams with no more than one loop and no more than three vertices. These are diagrams (2c) -- obtained by inserting two mass counterterms into (2a) --
\begin{equation}
    \Pi_{\mu\nu}^{(2c)}(p)=\frac{1}{2}\,m^{4}\frac{\partial^{2}}{\partial (m^{2})^{2}}\,\Pi_{\mu\nu}^{(2a)}(p)\ ,
\end{equation}
which is finite by power counting, and the tree-level mass counterterm on the top of Fig.~\ref{glu1ldiags}. The latter is easily seen to cancel the tree-level gluon mass introduced in the denominator of the dressed gluon propagator $\Delta_{\mu\nu}(p)$ by Eq.~\eqref{Sm}: suppressing the trivial color indices, one can express the dressed inverse propagator as
\begin{equation}
    \Delta_{\mu\nu}^{-1}(p)=(p^{2}+m^{2})\,t_{\mu\nu}(p)+\frac{p^{2}}{\xi}\,\ell_{\mu\nu}(p)-\Pi_{\mu\nu}(p)\ ,
\end{equation}
where $\Pi_{\mu\nu}(p)$, the gluon polarization tensor, can be split as
\begin{equation}
    \Pi_{\mu\nu}(p)=m^{2}\,t_{\mu\nu}(p)+\Pi_{\mu\nu}^{(\text{loops})}(p)\ ,
\end{equation}
$\Pi_{\mu\nu}^{(\text{loops})}(p)$ being the loops' contribution to the polarization. Clearly, then,
\begin{equation}\label{gldressed}
    \Delta_{\mu\nu}^{-1}(p)=p^{2}\,\left[t_{\mu\nu}(p)+\frac{1}{\xi}\,\ell_{\mu\nu}(p)\right]-\Pi_{\mu\nu}^{(\text{loops})}(p)\ ,
\end{equation}
where the first term in the above equation is the inverse bare gluon propagator of ordinary perturbation theory and, to one-loop, $\Pi_{\mu\nu}^{(\text{loops})}(p)$ is the sum of diagrams (1), (2a)-(2c) and (3a)-(3b). It follows from Eq.~\eqref{gldressed} that the zero-momentum limit of the gluon propagator is completely determined by the loops of the perturbative series. In particular, in the framework of the screened massive expansion, the saturation of the gluon propagator to a finite, non-zero value -- that is, mass generation in the gluon sector -- is dynamical in nature, in the sense that it is an effect of the radiative corrections to the propagator, and not of the mere shift of the perturbative series defined by Eq.~\eqref{Sm}.

The BRST invariance of the total Faddeev-Popov action can be used to prove that, in linear covariant gauges, both in vacuum and at finite temperature, the exact longitudinal component of the dressed gluon propagator is equal to its tree-level value -- namely,
\begin{equation}
    \Delta_{\ell}(p)=\ell^{\mu\nu}(p)\Delta_{\mu\nu}(p)=\frac{\xi}{p^{2}}\ .
\end{equation}
In the Landau gauge ($\xi=0$), in particular, $\Delta_{\ell}(p)$ vanishes. As for the transverse component of $\Delta_{\mu\nu}(p)$, the breaking of Euclidean O(4) symmetry down to O(3), caused by the existence, at $T\neq 0$, of a privileged direction $n^{\mu}=(1,0,0,0)$, gives rise to two distinct transverse sectors \cite{KG23}. The first one is obtained by projecting the propagator on the subspace which is transverse to both $p$ and $n_{\perp}(p)=n-(n\cdot p)\,p/p^{2}$ -- equivalently, to $p$ and $n$~--, which can be done using the projector $\mc{P}^{T}_{\mu\nu}(p)$,
\begin{equation}
    \mc{P}_{\mu\nu}^{T}(p)=(1-\delta_{\mu4})(1-\delta_{\nu4})\left(\delta_{\mu\nu}-\frac{p_{\mu}p_{\nu}}{{\bf p}^{2}}\right)\ ,
\end{equation}
where $p=(p^{4},{\bf p})=(\omega_{n},{\bf p})$. The second is obtained by projecting the propagator on the subspace which is transverse to $p$ but longitudinal to $n_{\perp}(p)$, defined by the projector $\mc{P}^{L}_{\mu\nu}(p)$,
\begin{equation}
    \mc{P}_{\mu\nu}^{L}(p)=t_{\mu\nu}(p)-\mc{P}_{\mu\nu}^{T}(p)\ .
\end{equation}
$\mc{P}^{T}_{\mu\nu}(p)$ and $\mc{P}^{L}_{\mu\nu}(p)$ satisfy the identities
\begin{align}
    p\cdot\mc{P}^{T,L}(p)=\mc{P}^{T,L}(p)\cdot p=0\ ,\\
    \notag n_{\perp}(p)\cdot\mc{P}^{T}(p)=\mc{P}^{T}(p)\cdot n_{\perp}(p)=0\ ,\\
    \notag n_{\perp}(p)\cdot\mc{P}^{L}(p)=\mc{P}^{L}(p)\cdot n_{\perp}(p)=n_{\perp}(p)\ ,\\
    \notag\mc{P}^{T,L}(p)\cdot \mc{P}^{T,L}(p) = \mc{P}^{T,L}(p)\ ,\\
    \notag t(p)\cdot \mc{P}^{T,L}(p)=\mc{P}^{T,L}(p)\cdot t(p)=\mc{P}^{T,L}(p)\ ,\\
    \notag \ell(p)\cdot \mc{P}^{T,L}(p)=\mc{P}^{T,L}(p)\cdot \ell(p)=0\ ,\\
    \notag\mc{P}^{T,L}(p)\cdot \mc{P}^{L,T}(p) = 0\ ,\\
    \text{Tr}\{\mc{P}^{T}(p)\}=2, \quad \text{Tr}\{\mc{P}^{L}(p)\}=1\ ,
\end{align}
where $\text{Tr}\{X\}=\delta^{\mu\nu}X_{\mu\nu}$ denotes the trace of $X$ over its spacetime indices.

Using the projectors $\mc{P}^{T,L}(p)$, the Landau-gauge gluon propagator at finite temperature can be expressed as
\begin{equation}
    \Delta_{\mu\nu}(p)=\Delta_{T}(\omega_{n},|{\bf p}|)\ \mc{P}^{T}_{\mu\nu}(p)+\Delta_{L}(\omega_{n},|{\bf p}|)\ \mc{P}^{L}_{\mu\nu}(p)\ ,
\end{equation}
where $\Delta_{T}(\omega_{n},|{\bf p}|)$ and $\Delta_{L}(\omega_{n},|{\bf p}|)$, evaluated as
\begin{align}
    \Delta_{T}(\omega_{n},|{\bf p}|)&=\frac{1}{2}\,\text{Tr}\{\mc{P}^{T}_{\mu\nu}(p)\,\Delta_{\mu\nu}(p)\}\ ,\\
    \Delta_{L}(\omega_{n},|{\bf p}|)&=\text{Tr}\{\mc{P}^{L}_{\mu\nu}(p)\,\Delta_{\mu\nu}(p)\}\ ,
\end{align}
are referred to, respectively, either as the (spatially) transverse and longitudinal propagators, or as the chromomagnetic and the chromoelectric propagator. Since in what follows we will only present results in the Landau gauge, for which $\Delta_{\ell}(p)=0$, from now on we will assume that the limit $\xi \to 0$ has been taken and we will use the terms ``transverse'' and ``longitudinal'' in the sense of $\mc{P}^{T,L}(p)$ defined above.

\begin{figure}[h]
    \centering
    \includegraphics[width=0.25\textwidth]{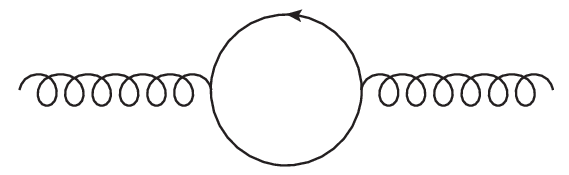}
    \caption{Full-QCD one-loop quark polarization diagram.}
    \label{gluqkdiag}
\end{figure}

Explicit expressions for the functions $\Delta_{T,L}(\omega_{n},|{\bf p}|)$ were derived to one loop in pure Yang-Mills theory in \cite{SC21}. Up to one-dimensional integrations weighted by the Bose distribution that can only be evaluated numerically, they are reported in analytic form. In order to obtain the corresponding expressions in full QCD, it is sufficient to add the ordinary one-loop quark diagram displayed in Fig.~\ref{gluqkdiag} to the polarization diagrams in Fig.~\ref{glu1ldiags}; just like the ghost loop (1) in the figure, the quark loop contains no internal gluon lines, and thus evaluates to its ordinary perturbative expression\footnote{Except for the fact that $M_{q}\sim\Lambda_{\text{QCD}}$, see the discussion in the previous section.}, reported in \cite{KG23,CS25} in terms of one-dimensional integrals of analytic functions weighted by the Fermi distribution.

In \cite{SC21,CS25}, the components $\Delta_{T,L}(\omega_{n}, |{\bf p}|)$ of the gluon propagator were expressed as
\begin{align}\label{ymglupropTraw}
    &\Delta_{T,L}^{-1}(\omega_{n},|{\bf p}|)=\\
    \notag&=p^{2}+\delta Z_{A}\,p^{2}-Ng^{2}\Pi_{T,L}^{(1)}(\omega_{n},|{\bf p}|)-g^{2}\sum_{f}\Pi_{T,L}^{(f)}(\omega_{n},|{\bf p}|)\ ,
\end{align}
where $p^{2}=\omega^{2}_{n}+{\bf p}^2$, $\delta Z_{A}$ is the gluon field-strength renormalization constant, $\Pi^{(1)}_{T,L}(\omega_{n},|{\bf p}|)$ -- modulo a factor of $Ng^{2}$ -- are the (diverging) transverse and longitudinal components of the gluon polarization terms computed from the loop diagrams in Fig.~\ref{glu1ldiags}, and $\Pi^{(f)}_{T,L}(\omega_{n},|{\bf p}|)$ -- modulo a factor of $g^{2}$ -- are those computed from the quark loop in Fig.~\ref{gluqkdiag}, one for each quark of flavor $f$. Of course, the latter must be omitted in pure Yang-Mills theory.

After renormalization, $\Delta_{T,L}(p)$ can be put in the form
\begin{align}\label{glupropadim}
    &p^{2}\Delta_{T,L}(p)=\\
    \notag&=\frac{z_{\pi}}{\pi_{1}(p)+[\pi_{T,L}(p)]_{\text{Th}}+\sum_{f}\{\pi_{f}(p)+[\pi_{T,L}^{(f)}(p)]_{\text{Th}}\}+\pi_{0}}\ ,
\end{align}
where $\pi_{0}$ is an additive renormalization constant, the coupling $\alpha_{s}$ was absorbed into the multiplicative factor $z_{\pi}$ and
\begin{align}
    \notag\pi_{1}(p)                       & =-\frac{16\pi^{2}}{3}\frac{[\Pi^{(1)}_{T,L}(p^{2})]_{\text{V,ren.}}}{p^{2}}\ , \\
    [\pi_{T,L}(p)]_{\text{Th}} & =-\frac{16\pi^{2}}{3}\frac{[\Pi^{(1)}_{T,L}(\omega_{n},|{\bf p}|)]_{\text{Th}}}{p^{2}}\ ,\\
    \notag\pi_{f}(p)                       & =-\frac{16\pi^{2}}{3N}\frac{[\Pi^{(f)}_{T,L}(p^{2})]_{\text{V,ren.}}}{p^{2}}\ , \\
    \notag[\pi_{T,L}^{(f)}(p)]_{\text{Th}} & =-\frac{16\pi^{2}}{3N}\frac{[\Pi^{(f)}_{T,L}(\omega_{n},|{\bf p}|)]_{\text{Th}}}{p^{2}}\ .
\end{align}
In the above expressions, the subscript ``V,ren.'' denotes renormalized vacuum polarization terms, whereas the subscript ``Th'' denotes purely thermal terms, which to one loop can be set apart from the former thanks to Eqs.~\eqref{msumb}, \eqref{msumf}.

The adimensional constants $z_{\pi}$ and $\pi_{0}$ in Eq.~\eqref{glupropadim}, together with the dimensionful value of the gluon mass parameter $m^{2}$, are the free parameters of the gluon propagator in the screened massive expansion. One of them~-- usually $z_{\pi}$ -- can be fixed in terms of the others by the renormalization conditions chosen for the propagator. This leaves us with two more free parameters, at variance with ordinary perturbation theory which only has one~-- namely, the coupling constant or, equivalently, the QCD scale $\Lambda_{\text{QCD}}$. In \cite{SC18} a method based on gauge invariance in linear covariant gauges was proposed to fix the value of $\pi_{0}$ at $T=0$ from first principles. There we found that requiring the phases of the residues of the gluon propagator to be independent from the gauge parameter $\xi$, in addition to the $\xi$-independence of the position of the corresponding poles -- which is a consequence of the so-called Nielsen identities \cite{Niel75a, Niel75b} --, yields a value $\pi_{0}=-0.876$ which is equal to that obtained by fitting the lattice data of \cite{DOS16} to within $\approx 1\%$. With $\pi_{0}$ fixed by principles of gauge invariance and $z_{\pi}$ fixed by renormalization, $m^{2}$ is the only remaining free parameter in the screened massive expansion of the gluon propagator. In this context, its role is similar to that of $\Lambda_{\text{QCD}}$ in ordinary perturbation theory, in that it sets the energy scale of the theory.

On the other hand, $m^{2}$ is introduced in the Lagrangian with the explicit aim of better approximating the order zero of perturbation theory. Therefore, when the expansion is extended to $T>0$, one should expect $m^{2}$ to be a function of temperature. Once $m^{2}=m^{2}(T)$, it is reasonable to expect that any optimization procedure employed to constrain the parameters of the screened massive expansion would lead to temperature-dependent results. In particular, we may assume that, for $T\geq 0$, $\pi_{0}=\pi_{0}(T)$ depends on temperature as well.

The optimization of the free parameters of the screened massive expansion by principles of gauge invariance was never extended to finite temperatures nor to full QCD\footnote{An early attempt to determine both the optimal value of the gluon propagator's $\pi_{0}$ constant and the mass scale generated in the quark sector by chiral symmetry breaking in vacuum proved unsuccessful, and the scope of the calculation was never narrowed to trying to obtain useful information in the gluon sector of full QCD alone.}. Because of this, in the following sections we will use two types of parameters to study the analytic structure of the gluon propagator. First, both at $T>0$ and in full QCD, we will use the same parameters as $T=0$ pure Yang-Mills theory optimized by gauge invariance. In addition to the aforementioned value of $\pi_{0}=-0.876$, we will take $m=0.656$~GeV, which in \cite{SC18} was found to accurately reproduce the lattice data of \cite{DOS16} for the vacuum gluon propagator. The second type of parameters we will employ were obtained by fitting lattice data for the static limit ($\omega_{n}=0$) of the propagator in pure Yang-Mills theory and in $n_{f}=2$ full QCD. As discussed in the previous section, in the static limit we find large differences in the parameters needed to fit the two components of the propagator. Since the transverse component shows a better agreement with the lattice data \cite{CS25,Com25b}, in the present study we will use the parameters obtained from the transverse sector. For more details on these parameters and on the data from which they were obtained, we refer to Secs.~III and IV below.

For all choices of parameters, we will impose on the gluon propagator spatial momentum-subtraction (MOM) renormalization conditions, defined by
\begin{equation}\label{rencond}
    \Delta_{T,L}(\omega_{n},|{\bf p}|)\Big|_{\omega_{n}=0,|{\bf p}|=\mu}=\frac{1}{\mu^{2}}\ .
\end{equation}
While $\Delta_{T}(p)\neq \Delta_{L}(p)$ for $T>0$, the two propagators are essentially indistinguishable from one another for $|{\bf p}|\gg T$, eventually approaching their $T\to 0$ limit as $|{\bf p}|\to \infty$. Therefore, as long as the renormalization scale $\mu$ is much larger than $T$ -- in what follows we will set $\mu=4$~GeV with $T\leq 700$~MeV --, Eq.~\eqref{rencond} effectively represents a unique renormalization condition for both the components -- equivalent to the MOM renormalization condition usually adopted in vacuum.

\subsection{Generalized spectral representation at finite temperature}

When analytically continued to complex frequencies $\omega_{n}\to z\in\mathbb{C}$, as we said, the one-loop gluon propagator computed in the screened massive expansion is found to have\footnote{Here we limit ourselves to discussing results obtained for parameter ranges compatible with optimization by gauge invariance and/or with lattice data. For more general results, see \cite{COM19}.} two symmetric pairs $\{z, -z,z^{\star},-z^{\star}\}$ of complex-conjugate poles ($\text{Re}\{z\},\text{Im}\{z\}\neq0$) in its principal Riemann sheet both in vacuum \cite{SIR16b,SIR17a,SIR17b,SC18,SC22a,SC22b,SC23} and at finite temperature \cite{SIR17d,SC21}, which translate to a single pair of complex-conjugate poles in the variable $z^{2}$. Because of these poles, the gluon propagator does not admit an ordinary K\"{a}ll\'{e}n-Lehmann spectral representation: if one could write \cite{KG23}
\begin{equation}\label{kallen}
\Delta_{T,L}(z,{\bf p})=\int_{-\infty}^{+\infty}d\omega\ \frac{\rho_{T,L}(\omega,{\bf p})}{\omega+iz}
\end{equation}
for some positive\footnote{It should be noted that, when entering the the K\"{a}ll\'{e}n-Lehmann representation of physical two-point functions as in the previous equation, the spectral densities are positive for $\omega>0$ and negative for $\omega<0$. For brevity, as in the introduction, when discussing positivity we will refer to the sign of the spectral function for $\omega>0$.} spectral density $\rho_{T,L}(\omega,{\bf p})$, then $\Delta_{T,L}(z,{\bf p})$'s poles in the variable $z\in\mathbb{C}$ would necessarily be pure imaginary -- that is, they would lie on the axis $p^{0}=ip^{4}=iz\in\mathbb{R}$, which is (naively) identified with the Minkowski axis\footnote{To expand on what we said in the introduction regarding the current understanding -- or lack thereof -- of complex-conjugate poles in the principal Riemann sheet of the propagator, whether this identification actually holds in the presence of such poles is disputed, see e.g. \cite{HK21b,SC22a}.}. Despite the lack of an ordinary K\"{a}ll\'{e}n-Lehmann representation, one can still leverage the analytic properties of the gluon propagator to derive a generalized spectral representation in which the complex poles' contribution appears alongside that of a spectral function defined, as usual, as the jump of the imaginary part of $\Delta_{T,L}(z,{\bf p})$ across its branch cut. While this generalized spectral representation as presented in \cite{SIR17a} only makes sense if the propagator is a function of $p^{2}$~-- that is, at $T=0$~--, it is easy to re-derive it in a slightly modified form which also applies to the case where the propagator is a function of $p^{4}=\omega_{n}$ and of $|{\bf p}|$ separately, thus making it suitable for calculations at finite temperature.\\

\begin{figure}[h]
    \centering
    \includegraphics[width=0.38\textwidth]{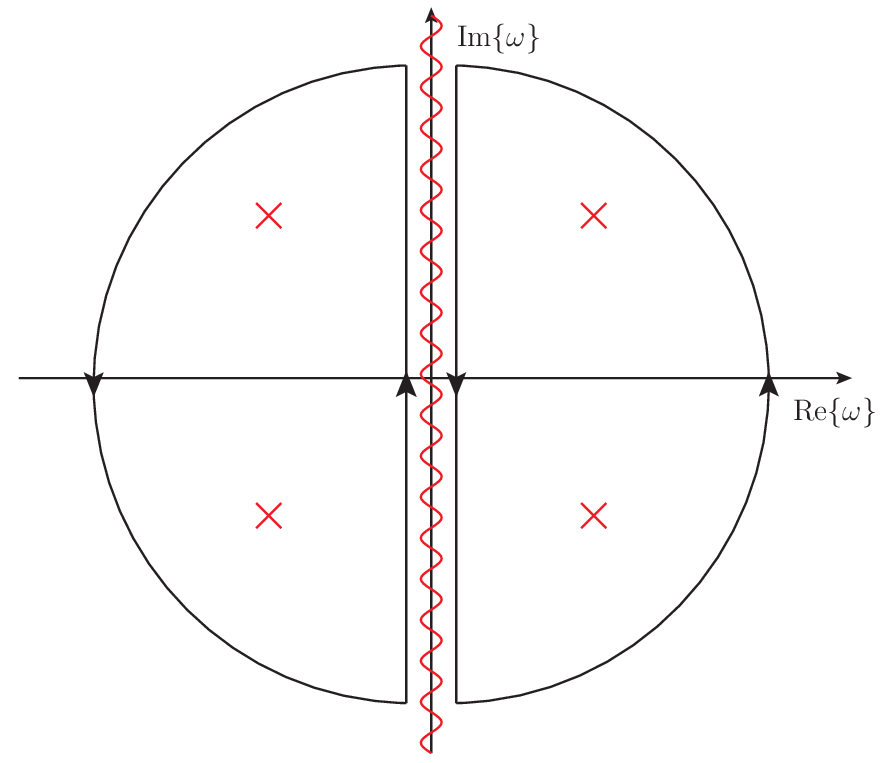}
    \caption{Integration contour used to derive the generalized spectral representation of the finite-temperature gluon propagator as a function of the spectral frequency $\omega$. The red crosses and wiggly line represent the singularities of the propagator: a finite number of complex poles in its principal Riemann sheet and a branch cut on the imaginary axis. The radius of the contour is sent to infinity.}
    \label{spectralcontour}
\end{figure}

Let us consider a complex function $\Delta(z)$, $z\in\mathbb{C}$, with the following properties: $\Delta(z)$
\begin{enumerate}[(i)]
\item goes to zero sufficiently fast as $|z|\to\infty$,
\item has a branch cut along the imaginary axis $i\mathbb{R}$,
\item has a finite number of poles $\{z_{k}\}$ of order one outside of the imaginary axis,
\item is analytic in $\mathbb{C}\setminus \left(\{z_{k}\}\cup i\mathbb{R}\right)$.
\end{enumerate}
Then, with $z\in\mathbb{C}\setminus \left(\{z_{k}\}\cup i\mathbb{R}\right)$, one can integrate $\Delta(w)/(w-z)$ along the contour shown in Fig.~\ref{spectralcontour} to obtain
\begin{equation}
    \oint\frac{dw}{2\pi i}\ \frac{\Delta(w)}{w-z}=\Delta(z)+\sum_{k}\frac{R_{k}}{z_{k}-z}\ ,
\end{equation}
where $R_{k}$ is the residue of $\Delta(z)$ at $z=z_{k}$. Since the integral along the contour at infinity vanishes, the left-hand side of the above equation also evaluates to
\begin{align}
    &\oint\frac{dw}{2\pi i}\ \frac{\Delta(w)}{w-z}=\\
    \notag&=-\int_{-i\infty+\epsilon}^{+i\infty+\epsilon}\frac{dw}{2\pi i}\frac{\Delta(w)}{w-z}+\int_{-i\infty-\epsilon}^{+i\infty-\epsilon}\frac{dw}{2\pi i}\frac{\Delta(w)}{w-z}=\\
    \notag&=\int_{-\infty}^{+\infty}\frac{d\omega}{2\pi}\frac{i\Delta(i\omega+\epsilon)}{\omega+iz}-\int_{-\infty}^{+\infty}\frac{d\omega}{2\pi}\frac{i\Delta(i\omega-\epsilon)}{\omega+iz}\ .
\end{align}
In particular, if we define a spectral function $\rho(\omega)$ as
\begin{equation}\label{spectr0}
    \rho(\omega)=\frac{-i}{2\pi}\,[\Delta(i\omega-\epsilon)-\Delta(i\omega+\epsilon)]\quad (\omega\in\mathbb{R}),
\end{equation}
we can then express $\Delta(z)$ as
\begin{equation}\label{genspectrprel}
    \Delta(z)=\sum_{k}\frac{R_{k}}{z-z_{k}}+\int_{-\infty}^{+\infty}d\omega\ \frac{\rho(\omega)}{\omega+iz}\ .
\end{equation}

From the above equation it follows that, if $\Delta(z)$ is real for $z\in\mathbb{R}$, then
\begin{enumerate}
    \item outside of the imaginary axis, $\Delta(z)$ can only have real poles\footnote{Note that, since $z\in\mathbb{R}$ here is the Euclidean axis, in the context of two-point Green functions of quantum field theories these are not the single-particle poles of physical propagators, but rather \textit{Euclidean} poles. Ordinary physical poles corresponding to asymptotic states are instead located on the imaginary axis $z\in i\mathbb{R}$ and, as we said in the introduction, they show up in spectral functions as Dirac deltas.} with real residues, or pairs of complex conjugate poles with complex conjugate residues,
    \item $[\rho(\omega)]^{\star}=-\rho(-\omega)$ for all $\omega\in\mathbb{R}$.
\end{enumerate}
If this is the case, then $[\Delta(z)]^{\star}=\Delta(z^{\star})$ for every $z\in\mathbb{C}$. Additionally, if $\Delta(z)$ is such that $\Delta(-z)=\Delta(z)$ for all $z\in\mathbb{R}$, then
\begin{enumerate}\setcounter{enumi}{2}
    \item for every pole $z_{k}$ with residue $R_{k}$, $-z_{k}$ is also a pole, with residue $-R_{k}$,
    \item $\rho(-\omega)=-\rho(\omega)$ for all $\omega\in\mathbb{R}$.
\end{enumerate}
If this is the case, then $\Delta(-z)=\Delta(z)$ for all $z\in\mathbb{C}$. In particular, if
\begin{equation}\label{propsymmetries}
    \Delta(z)^{\star}=\Delta(-z)=\Delta(z)
\end{equation}
for all $z\in\mathbb{R}$, we can use the chain of identities
\begin{align}
\Delta(i\omega+\epsilon)=\Delta(-i\omega+\epsilon)^{\star}=\Delta(i\omega-\epsilon)^{\star}
\end{align}
$(\omega\in\mathbb{R})$ to express the spectral function in Eq.~\eqref{spectr0} as
\begin{equation}
    \rho(\omega)=\frac{1}{\pi}\,\text{Im}\{\Delta(i\omega-\epsilon)\}\quad(\omega\in\mathbb{R})\ .
\end{equation}
The above equation shows that, for functions $\Delta(z)$ with the property in Eq.~\eqref{propsymmetries}, the spectral function $\rho(\omega)$ in Eq.~\eqref{genspectrprel} is a real function and -- modulo prefactors~-- it is equal to the jump of the imaginary part of $\Delta(z)$ across its discontinuity on the imaginary axis.\\

At finite temperature, the components $\Delta_{T,L}(\omega,|{\bf p}|)$ of the Euclidean gluon propagator computed in the screened massive expansion \cite{SC21,CS25} -- when analytically continued to the complex plane $\omega\to z\in\mathbb{C}$ -- satisfy properties (i) to (iv) for all fixed spatial momenta $|{\bf p}|$; additionally, $\Delta_{T,L}(\omega,|{\bf p}|)^{\star}=\Delta_{T,L}(-\omega,|{\bf p}|)=\Delta_{T,L}(\omega,|{\bf p}|)$ for all $\omega\in\mathbb{R}$. Therefore we can write
\begin{equation}\label{genspectr}
    \Delta_{T,L}(z, |{\bf p}|)=\sum_{k}\frac{R_{k}^{{T,L}}(|{\bf p}|)}{z-z_{k}^{{T,L}}(|{\bf p}|)}+\int_{-\infty}^{+\infty}d\omega\ \frac{\rho_{T,L}(\omega,|{\bf p}|)}{\omega+iz}\ ,
\end{equation}
where $\rho_{T,L}(\omega,|{\bf p}|)$,
\begin{equation}\label{specfun}
    \rho_{T,L}(\omega,|{\bf p}|)=\frac{1}{\pi}\,\text{Im}\{\Delta_{T,L}(i\omega-\epsilon,|{\bf p}|)\}\quad(\omega\in\mathbb{R})
\end{equation}
are momentum-dependent spectral functions. In the absence of complex conjugate poles, assuming the spectral functions are positive definite, Eq.~\eqref{genspectr} has the same form as the K\"{a}ll\'{e}n-Lehmann representation of a physical two-point function at finite temperature, Eq.~\eqref{kallen}. Because of the different procedure (and assumptions) followed to derive it, however, not only Eq.~\eqref{genspectr} is allowed to contain pole contributions, but it also implies no positivity constraint on the spectral functions. And indeed, as we will see in the following sections, just like in the vacuum, the spectral function of the gluon propagator at $|{\bf p}|=0$ and $T\neq0$ is not positive definite.

The generalized spectral representation of \cite{SIR17a} can be explicitly recovered in the limit $T\to 0$ as follows. First of all, we observe that, as soon as O(4) invariance is restored, with a slight abuse of notation and suppressing the $T,L$ labels (irrelevant in vacuum),
\begin{align}
    \rho(\omega,|{\bf p}|)&=\frac{1}{\pi}\,\text{Im}\{\Delta(i\omega-\epsilon,|{\bf p}|)\}=\\
    \notag&=\frac{1}{\pi}\,\text{Im}\{\Delta((i\omega-\epsilon)^{2}+{\bf p}^{2})\}=\\
    \notag&=\frac{1}{\pi}\,\text{Im}\{\Delta(-\omega^{2}+{\bf p}^{2}-i\epsilon\,\text{sgn}(\omega))\}=\\
    \notag&=\text{sgn}(\omega)\,\rho_{\text{vac}}(\mu^{2})|_{\mu^{2}=\omega^{2}-{\bf p}^{2}}\ ,
\end{align}
where $\rho_{\text{vac}}(\mu^{2})$ is the ordinary, vacuum spectral function. It follows that
\begin{align}
    &\int_{-\infty}^{+\infty}d\omega\ \frac{\rho(\omega,|{\bf p}|)}{\omega+iz}=\\
    \notag&=\int_{0}^{+\infty}d\omega\ \left\{\frac{\rho(\omega,|{\bf p}|)}{\omega+iz}+\frac{\rho(-\omega,|{\bf p}|)}{-\omega+iz}\right\}=\\
    \notag&=\int_{0}^{+\infty}d\omega\ \rho_{\text{vac}}(\omega^{2}-{\bf p}^{2})\ \left\{\frac{1}{\omega+iz}-\frac{1}{-\omega+iz}\right\}=\\
    \notag&=\int_{0}^{+\infty}d\mu^{2}\ \frac{\rho_{\text{vac}}(\mu^{2}-{\bf p}^{2})}{\mu^{2}+z^{2}}=\\
    \notag&=\int_{0}^{+\infty}d\mu^{2}\ \frac{\rho_{\text{vac}}(\mu^{2})}{\mu^{2}+z^{2}+{\bf p}^{2}}\ ,
\end{align}
where we have used the fact that, since on the Euclidean axis the propagator is real and has no discontinuity, $\rho_{\text{vac}}(\mu^{2})=\text{Im}\{\Delta(-\mu^{2}-i\epsilon)\}/\pi$ vanishes for $\mu^{2}<0$. Additionally, since every pole $z_{k}$ with residue $R_{k}$ in Eq.~\eqref{genspectr} appears together with the opposite pole $-z_{k}$ with residue $-R_{k}$,
\begin{align}
    \sum_{k}\frac{R_{k}(|{\bf p}|)}{z-z_{k}(|{\bf p}|)}=\sum_{k}{}^{\prime}\ \frac{2z_{k}(|{\bf p}|)R_{k}(|{\bf p}|)}{z^{2}-z_{k}^{2}(|{\bf p}|)}\ ,
\end{align}
where the sum on the right-hand side extends to half of the original poles. By O(4) invariance, when computed for $z=\omega\in\mathbb{R}$, the right-hand side of the last equation can only depend on the combination $\omega^{2}+{\bf p}^{2}$. Hence one must have 
\begin{equation}
    z_{k}(|{\bf p}|)=i\varepsilon_{k}(|{\bf p}|)=i\sqrt{m_{k}^{2}+{\bf p}^{2}}
\end{equation}
for some $|{\bf p}|$-independent complex constant $m_{k}^{2}$. For the same reason, $2z_{k}(|{\bf p}|)R_{k}(|{\bf p}|)$ must be independent of $|{\bf p}|$: it must be $2z_{k}(|{\bf p}|)R_{k}(|{\bf p}|)=\mc{R}_{k}$ for some complex constant $\mc{R}_{k}$. In conclusion, with $p^{2}=z^{2}+{\bf p}^{2}$, at $T=0$,
\begin{equation}
    \Delta(z, {\bf p})=\sum_{k}{}^{\prime}\frac{\mathcal{R}_{k}}{p^{2}+m^{2}_{k}}+\int_{0}^{+\infty}d\mu^{2}\ \frac{\rho_{\text{vac}}(\mu^{2})}{\mu^{2}+p^{2}}\ ,
\end{equation}
which is the generalized spectral representation of \cite{SIR17a}.

\section{Poles and spectral function of the gluon propagator in pure Yang-Mills theory}

In the present section we investigate the poles and the spectral function of the Landau-gauge gluon propagator at vanishing spatial momentum $|{\bf p}|=0$, as computed to one loop at finite temperature in the framework of the screened expansion of pure Yang-Mills theory. As anticipated in Sec.~IIB, we will present results for two sets of parameters. In Sec.~IIIA we fix $\pi_{0}=-0.876$ and $m=0.656$~GeV. These values were obtained at $T=0$, respectively, by optimizing the gluon propagator by principles of gauge invariance -- see \cite{SC18} and the discussion in Secs.~IIA-B -- and by fitting the gluon mass parameter $m^{2}$ to the lattice data of \cite{DOS16} given said value of $\pi_{0}$. In Sec.~IIIB we use temperature-dependent parameters $\pi_{0}(T)$ and $m^{2}(T)$ obtained by fitting the finite-temperature lattice data of \cite{SOBC14} for the Landau-gauge chromomagnetic gluon propagator $\Delta_{T}(\omega_{n},|{\bf p}|)$ at zero Matsubara frequency $\omega_{n}=0$. Both sets of parameters were already used in \cite{CS25} to study the properties of the gluon propagator in the static limit.

\subsection{Optimized vacuum parameters}

\begin{figure}[h]
    \centering
    \includegraphics[width=0.33\textwidth,angle=270]{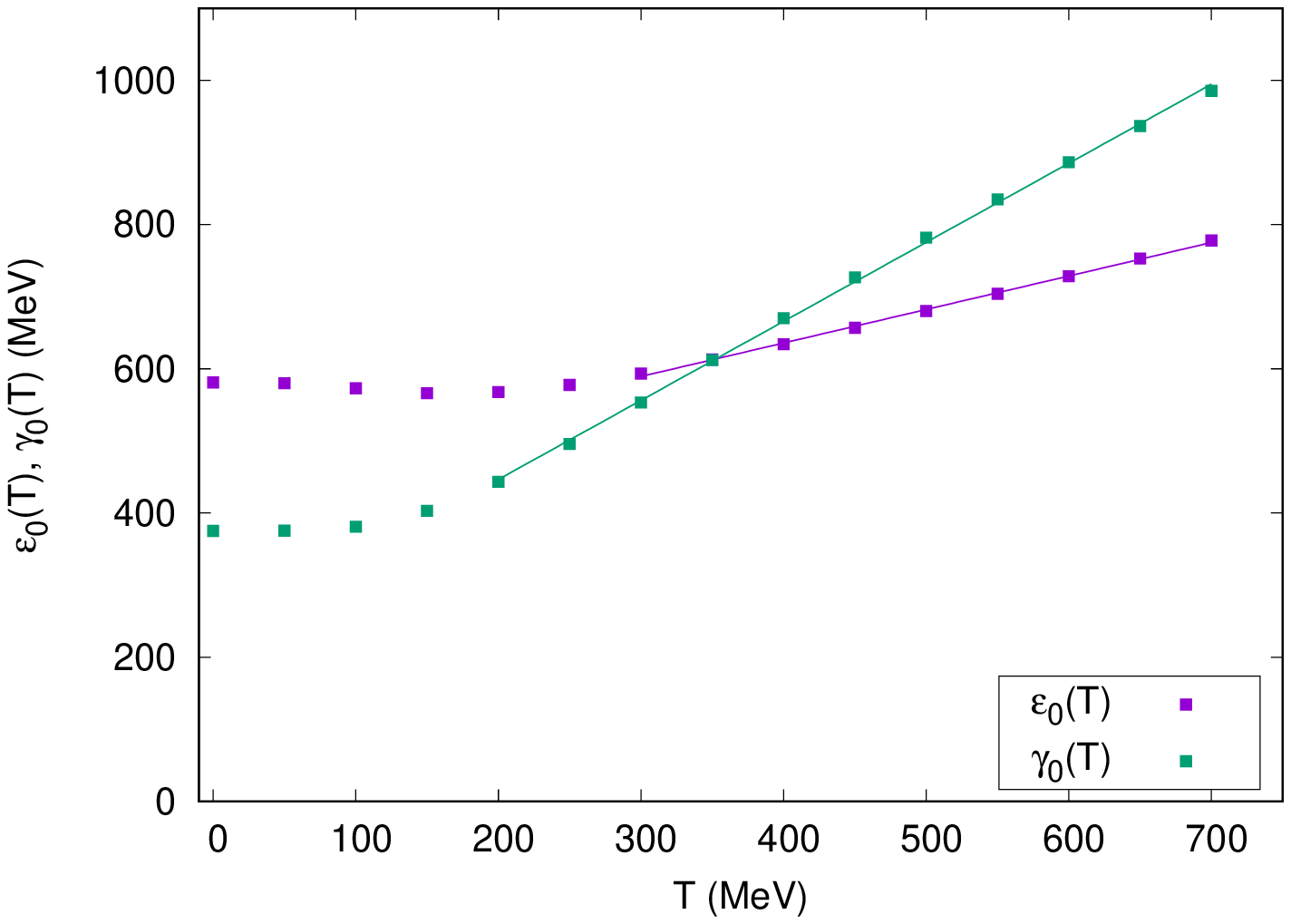}
    \caption{Real and imaginary part of one of the four symmetric, complex-conjugate poles $z(T)=i[\varepsilon_{0}(T)+i\gamma_{0}(T)]$ of the Euclidean Landau-gauge pure-Yang-Mills gluon propagator at zero spatial momentum $|{\bf p}|=0$, as functions of the temperature $T$. Computed for $m=0.656$~GeV, $\pi_{0}=-0.876$.}
    \label{poles_ym_fixed}
\end{figure}

The zero-momentum ($|{\bf p}|=0$) poles of the Landau-gauge gluon propagator were first computed at $T\geq 0$ using temperature-independent parameters\footnote{Ref.~\cite{SIR17d} predates the optimization of the screened massive expansion and used different values of $m$ and $\pi_{0}$. There, like in other references, $\pi_{0}$ was denoted $F_{0}$.} in \cite{SIR17d}. In Fig.~\ref{poles_ym_fixed} we display them for $m=0.656$~GeV, $\pi_{0}=-0.876$, as a function of temperature. As discussed in Sec.~IIC, because of the property $\Delta_{T,L}(\omega,|{\bf p}|)^{\star}=\Delta_{T,L}(-\omega,|{\bf p}|)=\Delta_{T,L}(\omega,|{\bf p}|)$, $\omega\in\mathbb{R}$, the poles come in symmetric, complex-conjugate quartets $\{z, -z, z^{\star}, -z^{\star}\}$, where $z$ is the complexified frequency, on the principal Riemann sheet. Just like in vacuum \cite{SIR16b,SC18}, at finite temperature we find one such set of poles. In the figure, we set $z=i(\varepsilon_{0}+i\gamma_{0})$ and only show the pole with $\varepsilon_{0},\gamma_{0}>0$. Following the Minkowski-space point of view -- by which, for a physical pole, $\varepsilon_{0}$ and $\gamma_{0}$ would be, respectively, the zero-momentum energy (mass) and damping factor -- we will call $\varepsilon_{0}$ and $\gamma_{0}$ the real and the imaginary part of the pole, respectively.

When computed at fixed, optimized $T=0$ parameters, the real part of the pole $\varepsilon_{0}(T)$ slightly decreases from its vacuum value of $\varepsilon_{0}(T=0)=581$~MeV \cite{SC18} down to $\approx566$~MeV at $T\approx165$~MeV, then starts to increase at larger temperatures. For $T\gtrapprox300$~MeV, its behavior is essentially linear with $T$. A fit of $\varepsilon_{0}(T)$ at temperatures between $300$ and $700$ MeV yields
\begin{equation}
    \varepsilon_{0}(T)\approx 449.9\text{ MeV} + 0.4645 \,T\ \quad(T\in[300,700]\text{ MeV})\ .
\end{equation}
The imaginary part of the pole $\gamma_{0}(T)$, on the other hand, strictly increases with $T$, with a linear behavior setting in at a lower temperature than that of $\varepsilon_{0}(T)$. At $T=0$ it attains the value $\gamma_{0}(T=0)=375$~MeV \cite{SC18}, then it starts to increase linearly at $T\approx200$~MeV, until it becomes equal in magnitude to $\varepsilon_{0}(T)$ around $T\approx350$~MeV. At higher temperatures, $\gamma_{0}(T)>\varepsilon_{0}(T)$. For $T\in[200,700]$~MeV, $\gamma_{0}(T)$ is well approximated by the linear function
\begin{equation}\label{ymfixedlinearg}
    \gamma_{0}(T)\approx227.2\text{ MeV}+1.0963\,T\ \quad(T\in[200,700]\text{ MeV})\ .
\end{equation}

\begin{figure}[h]
    \centering
    \includegraphics[width=0.33\textwidth,angle=270]{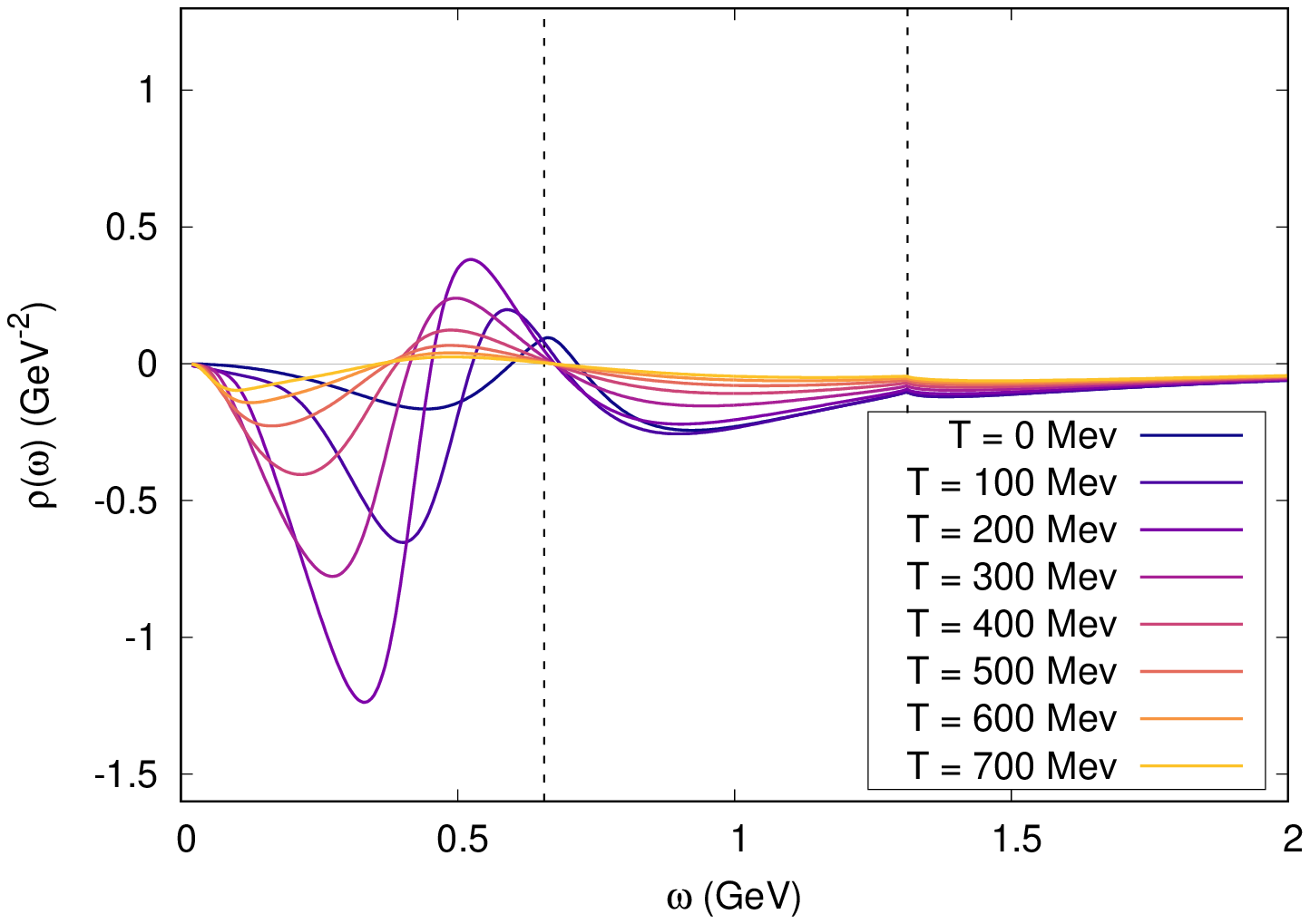}
    \caption{Zero-momentum ($|{\bf p}|=0$) spectral function of the Landau-gauge pure-Yang-Mills gluon propagator as a function of the spectral frequency $\omega$, for different values of the temperature $T$. The dashed vertical lines mark the $m$ and $2m$ mass thresholds. Computed for $m=0.656$~GeV, $\pi_{0}=-0.876$.}
    \label{spectral_ym_fixed}
\end{figure}

In Fig.~\ref{spectral_ym_fixed} we display the zero-momentum spectral function $\rho(\omega)=\rho_{T,L}(\omega,|{\bf p}|=0)$ -- see Eq.~\eqref{specfun} -- as a function of the spectral frequency $\omega>0$ for different values of the temperature $T$. By the properties of the propagator, $\rho(-\omega)=-\rho(\omega)$, so we won't show the $\omega<0$ portion of the curve.

For $T=0$ and $\omega>0$, $\rho_{T,L}(\omega,0)=\rho_{\text{vac}}(p^{2})|_{p^{2}=\omega^{2}}$, where $\rho_{\text{vac}}(p^{2})$ is the ordinary vacuum spectral function, already presented in Ref.~\cite{SC18} for the same values of the parameters and before that in \cite{SIR16b} for different values. There it was found that $\rho_{\text{vac}}(p^{2})$ is not positive definite, but is instead negative at large values of $p^{2}$ -- a feature that, as we said in the introduction, immediately follows from asymptotic freedom and from the negativity of the propagator's anomalous dimension in the UV \cite{OZ80}. At finite temperature and for large $\omega$, $\rho(\omega)$ is still negative. At fixed $\omega$, it flattens to zero from below as $T$ increases. It does not become uniformly positive above some value of the temperature: in the asymptotic limit $\omega \to \infty$, given that trivially $\omega\gg T$, thermal corrections become negligible and the spectral function approaches its negative vacuum limit.

At all temperatures, after attaining a first local minimum for $\omega>2m\approx1.312$~GeV, $\rho(\omega)$ shows a cusp at $\omega=2m$ -- marked in the figure with a dashed vertical line. This is a direct effect of the structure of the diagrams contributing to the gluon polarization to one loop. In more detail, diagram (3a) in Fig.~\ref{glu1ldiags} contains two internal gluon lines $1/(k^{2}+m^{2})$, each of which is multiplied by a transverse projector $t_{\mu\nu}(k)=\delta_{\mu\nu}-k_{\mu}k_{\nu}/k^{2}$ bearing a factor of $1/k^{2}$. Since
\begin{equation}
\frac{1}{k^{2}(k^{2}+m^{2})}=\frac{1}{m^{2}}\left(\frac{1}{k^{2}}-\frac{1}{k^{2}+m^{2}}\right)\ ,
\end{equation}
the diagram can be organized into three types of terms, characterized by how many massive denominators they contain -- two, one or zero. All such terms have singularities for $p^{2}\leq0$: those with two massive denominators are singular for $|p^{2}| \geq 4m^{2}$, those with one are singular for $|p^{2}|\geq m^{2}$, whereas those with no massive denominator are singular for all $|p^{2}|$. Similarly, diagram (1) in Fig.~\ref{glu1ldiags} contains two massless ghost lines and is thus singular for all $p^{2}\leq 0$. Diagram (3b) can be obtained by differentiating (3a), so it shares the same singularities, whereas the tadpoles (2a)-(2c) evaluate to constants, and are thus trivially non-singular.

The pole singularities described above, when integrated, translate to logarithmic singularities on the imaginary axis of the complexified variable $\omega$ of the propagator $\Delta_{T,L}(\omega,|{\bf p}|)$, which are precisely the branch cuts used in Sec.~IIC to derive its generalized spectral representation. At $|{\bf p}|=0$, in terms of its variable $\omega$, they yield contributions to the spectral function for $|\omega|\geq2m$, $|\omega|\geq m$ and $|\omega|\geq 0$ (that is, all $\omega$), the first two of which vanish identically below the corresponding thresholds, thus producing non-analyticities in $\rho(\omega)$. The cusp in Fig.~\ref{spectral_ym_fixed}~-- showing a discontinuity in the derivative $\partial\rho/\partial\omega$ at the two-particle threshold $\omega=2m$ -- is the expression of this non-analytic behavior. Conversely, no discontinuity is observed in the derivative of $\rho(\omega)$ at $\omega=m=0.656$~MeV, also marked by a dashed line in the figure. Upon inspection of the semi-analytic expressions in \cite{SC21} and of their analytic $T\to0$ limit, the reason for this can be traced back to extra factors of $p^{2}+m^{2}$ multiplying all integrals with only one massive denominator, which make the corresponding contributions vanish in the $p^{2}\to -m^{2}$~-- equivalently, $|\omega|\to m$ -- limit.

Between $\omega=m$ and $\omega=2m$, the spectral function attains a second local minimum\footnote{Albeit not so clear from the figure, the presence of this minimum at higher temperatures can be verified numerically.} at a position which first decreases from $\omega\approx0.93$~GeV ($T=0$) to $\omega\approx0.89$~GeV ($T\approx150$~MeV), and then shifts towards $\omega=2m$ with temperature for $T\gtrapprox150$~MeV. The singularity at $\omega=m$ is then replaced with a temperature-dependent maximum located at $\omega \sim m$; it is in a neighborhood of this maximum that $\rho(\omega)$ becomes positive in an interval that widens as the temperature increases, while the position of the maximum moves towards lower $\omega$'s. Finally, before vanishing at $\omega=0$, the spectral function attains a third local minimum at a position that decreases with temperature. The behavior of this minimum is specular to that of its preceding maximum, in that both the depth of the former and the height of the latter are non-monotonic with temperature, increasing as $T$ grows from zero to $T\approx190$~MeV and then flattening down at higher temperatures.

\subsection{Temperature-dependent lattice parameters}

\begin{figure}[h]
    \centering
    \includegraphics[width=0.33\textwidth,angle=270]{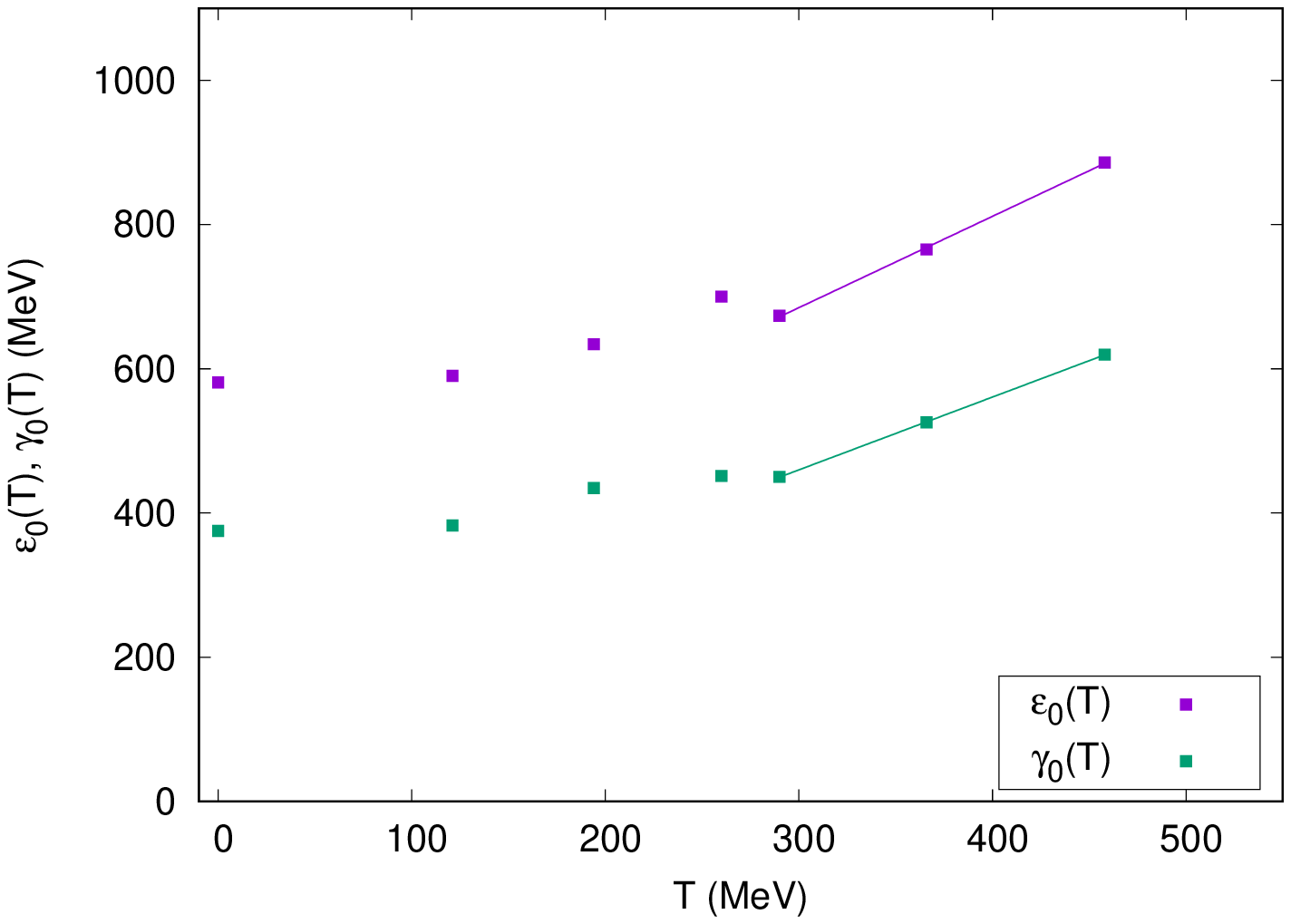}
    \caption{Real and imaginary part of one of the four symmetric, complex-conjugate poles $z(T)=i[\varepsilon_{0}(T)+i\gamma_{0}(T)]$ of the Euclidean Landau-gauge pure-Yang-Mills gluon propagator at zero spatial momentum $|{\bf p}|=0$, as functions of the temperature $T$. Computed using temperature-dependent $m^{2}=m^{2}(T)$, $\pi_{0}=\pi_{0}(T)$, as reported in Tab.~\ref{ym_lattice_params}.}
    \label{poles_ym_lattice}
\end{figure}

In Fig.~\ref{poles_ym_lattice} we display the zero-momentum gluon poles computed by making use of the parameters reported in Tab.~\ref{ym_lattice_params}. These were obtained in \cite{CS25} by fitting the lattice data of \cite{SOBC14} for the temperature-dependent transverse component of the Landau-gauge gluon propagator at zero Matsubara frequency ($\omega_{n}=0$). The point at $T=0$ in the figure is the optimized vacuum fit of the previous section, not part of the data of \cite{SOBC14}. A comparison with the lowest-temperature data point in the table, $T=121$~MeV, suggests that the behavior of pure Yang-Mills theory at $T\lessapprox100$~MeV is well approximated by that of vacuum.

\begin{table}
    \setlength{\tabcolsep}{9pt}
    \begin{tabular}{|c|c|c|}
        \hline
        $T$ (MeV) & $m(T)$ (MeV) & $\pi_{0}(T)$ \\
        \hline
        121 & 675 & -0.83\\
        194 & 725 & -0.78\\
        260 & 775 & -0.58\\
        290 & 725 & -0.48\\
        366 & 800 & -0.38\\
        458 & 900 & -0.28\\
        \hline
    \end{tabular}
    \caption{Values of the parameters $m(T)$ and $\pi_{0}(T)$ obtained in \cite{CS25} by fitting the lattice data of \cite{SOBC14} for the transverse component of the Landau-gauge Euclidean gluon propagator at zero Matsubara frequency ($\omega_{n}=0$), renormalized at $\mu=4$~GeV in the MOM scheme (see Sec.~IIB).}
    \label{ym_lattice_params}
\end{table}

As we can see from the figure, the evolution of $\varepsilon_{0}(T)$ and $\gamma_{0}(T)$ with $T$ computed by letting the parameters depend on temperature is quite different from the one described in the previous section. The real part of the pole does not show an initial decrease with temperature, but monotonically increases instead\footnote{The small decrease between $T=260$~MeV and $T=290$~MeV is the direct consequence of a corresponding drop in the mass parameter $m(T)$ which is within the indeterminacy of the fit. We are not able to confirm its significance.}. If we assume that the $T\to 0$ limit of the present dataset is consistent with the point at $T=0$ of \cite{DOS16}, then $\varepsilon_{0}(T)$ displays a plateau at low temperatures and starts increasing at $T\gtrapprox100$~MeV. Above $T= T_{c}\approx270$~MeV~-- the critical temperature of the deconfinement transition of pure Yang-Mills theory \cite{SOBC14} --, $\varepsilon_{0}(T)$ becomes again essentially linear with temperature, with an intercept and a slope given by
\begin{equation}\label{ym_latt_pole_re_fit}
    \varepsilon_{0}(T)\approx304.9\text{ MeV} + 1.2662\,T\quad(T\in[290,458]\text{ MeV})\ .
\end{equation}
At $T=458$~MeV, the highest temperature fitted from the lattice, the real part of the pole is sensibly larger than when computed by optimized vacuum parameters: we find $\varepsilon_{0}(T=458\text{ MeV})=886$~MeV, against $657$~MeV at $T=450$~MeV in Sec.~IIIA -- see Fig.~\ref{poles_ym_fixed}.

One of the most interesting features of Fig.~\ref{poles_ym_lattice} is the absence of crossing between the real and the imaginary part of the gluon pole which we observe in fixed-parameter calculations. While the imaginary part $\gamma_{0}(T)$ still increases monotonically with temperature, the increase computed with the parameters in Tab.~\ref{ym_lattice_params} is milder and keeps $\gamma_{0}(T)<\varepsilon_{0}(T)$ across the considered temperature range. At $T=458$~MeV, $\gamma_{0}(T)=620$~MeV, which is roughly $100$~MeV smaller than the value $\gamma_{0}(T=450\text{ MeV})=727$~MeV calculated at fixed vacuum parameters.

Another noteworthy aspect of the poles evaluated with lattice parameters is the fact that the linear behavior of the real and imaginary part now seems to set in at approximately the same temperature -- namely, the deconfinement temperature $T_{c}$~-- instead of being slightly offset as in Sec.~IIIA. For $T> T_{c}$, $\gamma_{0}(T)$ is well approximated by the function
\begin{equation}
    \gamma_{0}(T)\approx157.0\text{ MeV} + 1.0094\,T\quad(T\in[290,458]\text{ MeV})\ .
\end{equation}
We should note, however, that both the last equation and Eq.~\eqref{ym_latt_pole_re_fit} were obtained by considering only three temperature data points. We cannot thus exclude that the high-temperature linear behavior observed in Fig.~\ref{poles_ym_lattice} for both $\varepsilon_{0}(T)$ and $\gamma_{0}(T)$ is just an approximation of a non-linear one.
\begin{figure}[h]
    \centering
    \includegraphics[width=0.33\textwidth,angle=270]{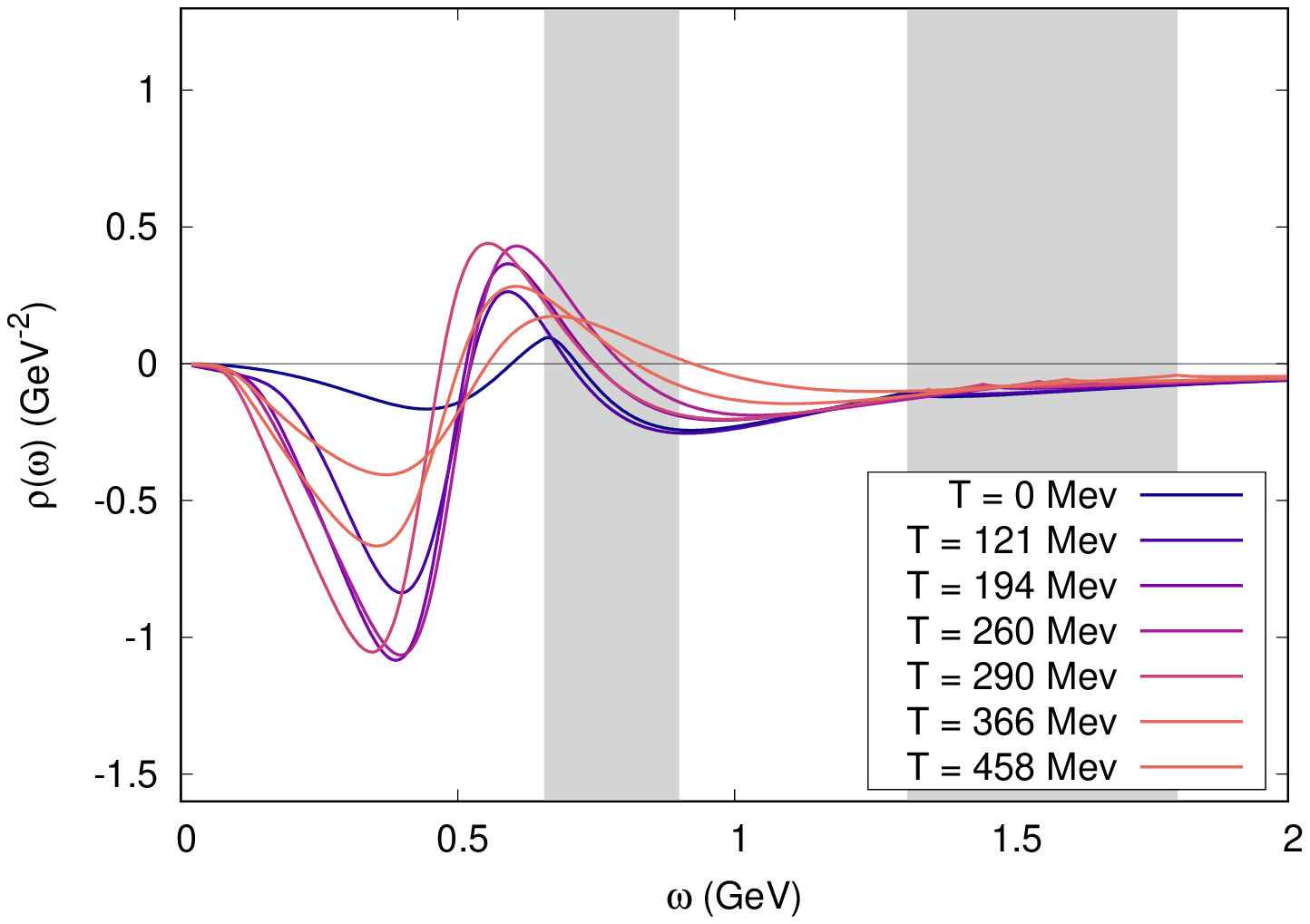}
    \caption{Zero-momentum ($|{\bf p}|=0$) spectral function of the Landau-gauge pure-Yang-Mills gluon propagator as a function of the spectral frequency $\omega$, for different values of the temperature $T$. Computed using temperature-dependent $m^{2}=m^{2}(T)$, $\pi_{0}=\pi_{0}(T)$, as reported in Tab.~\ref{ym_lattice_params}. The grey bands mark the regions $\omega\in[m_{\text{min}},m_{\text{max}}]$ and $\omega\in[2m_{\text{min}},2m_{\text{max}}]$, where $m_{\text{min,max}}$ are the minimum and maximum gluon mass parameter in the table.}
    \label{spectral_ym_lattice}
\end{figure}

In Fig.~\ref{spectral_ym_lattice} we display the zero-momentum gluon spectral function $\rho(\omega)=\rho_{T,L}(\omega,|{\bf p}|=0)$ as a function of $\omega\geq0$ for different values of temperature, using the same parameters as Fig.~\ref{poles_ym_lattice}. Like in the latter, the $T=0$ curve computed using the optimized vacuum parameters is added for reference.

The $T>0$ spectral function computed using parameters from the lattice shows a behavior similar to that described in Sec.~IIIA, with a few differences. First of all, once the gluon mass parameter is made to depend on the temperature $T$, the two-particle threshold $\omega=2m$ becomes $T$-dependent as well, shifting to higher temperatures together with the discontinuity in $\partial\rho/\partial \omega$ as indicated by the values of $m$ in Tab.~\ref{ym_lattice_params}. Second, both the positive maxima and the lowest-lying minima, which in Fig.~\ref{spectral_ym_fixed} showed a clear dependence on temperature, display a tendency to line up. For $T$ between $121$~MeV and $366$~MeV, the maxima are found at a position $\omega\in[0.55,0.61]$~GeV, whereas for $T=458$~MeV the maximum is found at a slightly higher frequency, $\omega\approx0.68$~GeV. Similarly, for all considered temperatures, the lowest-lying minimum is found between $\omega=0.34$~GeV and $\omega=0.40$~GeV. This behavior reflects the increase of the gluon mass parameter $m^{2}(T)$ with temperature, which compensates the shift of the position of the extrema towards smaller $\omega$'s observed in Sec.~IIIA. Finally, the lowest-lying minimum and the positive maximum are, respectively, deepest and highest at temperatures in the range $T\in[194,290]$~MeV and $T\in[260,290]$~MeV, both of which include the deconfinement temperature $T_{c}\approx 270$~MeV.

\section{Poles and spectral function of the gluon propagator in full QCD}

In this section we investigate the poles and the spectral function of the Landau-gauge gluon propagator at vanishing spatial momentum $|{\bf p}|=0$, as computed to one loop at finite temperature and zero quark density in the framework of the screened massive expansion of full QCD using the model discussed in Sec.~IIA for the quark masses. Like we did in Sec.~III, we will first present results obtained by using the optimized parameters derived in \cite{SC18} within $T=0$ pure Yang-Mills theory. For this calculation we will choose the same $n_{f}=2+1$ quark configuration employed in \cite{CS25} to study the deconfinement phase transition. Following that, in Sec.~IVB, we will display the poles and spectral function computed using temperature-dependent parameters obtained by fitting \cite{Com25b} recently reported unquenched lattice data for the static limit ($\omega_{n}=0$) of the Landau-gauge transverse gluon propagator in $n_{f}=2$ QCD \cite{SOS25}. Regardless of temperature, the fits were found to be largely insensitive to the quark masses $M_{q}$ over a wide range of values, going from $M_{q}=5$~MeV to $M_{q}=800$~MeV: such a large variation could always be compensated by a change of less than $50$~MeV in the gluon mass parameter $m$ and a modest change in $\pi_{0}$, without meaningfully altering the quality of the fits. In Sec.~IVB we will fix $M_{q}=400$~MeV, which is close to the infrared mass generated by chiral symmetry breaking for light quarks in vacuum \cite{KBLW05,CRBS21}.

As discussed in Sec.~IIA, the quark mass scale is expected to depend on temperature due to both chiral symmetry restoration and ordinary perturbative thermal effects. Fully accounting for such effects is currently out of our reach, as even in our simplified treatment of the quark masses it would require a self-consistent knowledge of the variation of $M_{q}$ with $T$ within the framework of the screened massive expansion, which at this time is unavailable. Still, an indication on how the analytic structure of the screened expansion's one-loop Landau-gauge gluon propagator may change with the quark mass scale can be obtained by an explicit calculation using different values of $M_{q}$. Sec.~IVC reports this calculation, both for $n_{f}=2+1$ fixed, optimized parameters, and for $n_{f}=2$ lattice parameters.

\subsection{Optimized vacuum parameters}

\begin{figure}[h]
    \centering
    \includegraphics[width=0.33\textwidth,angle=270]{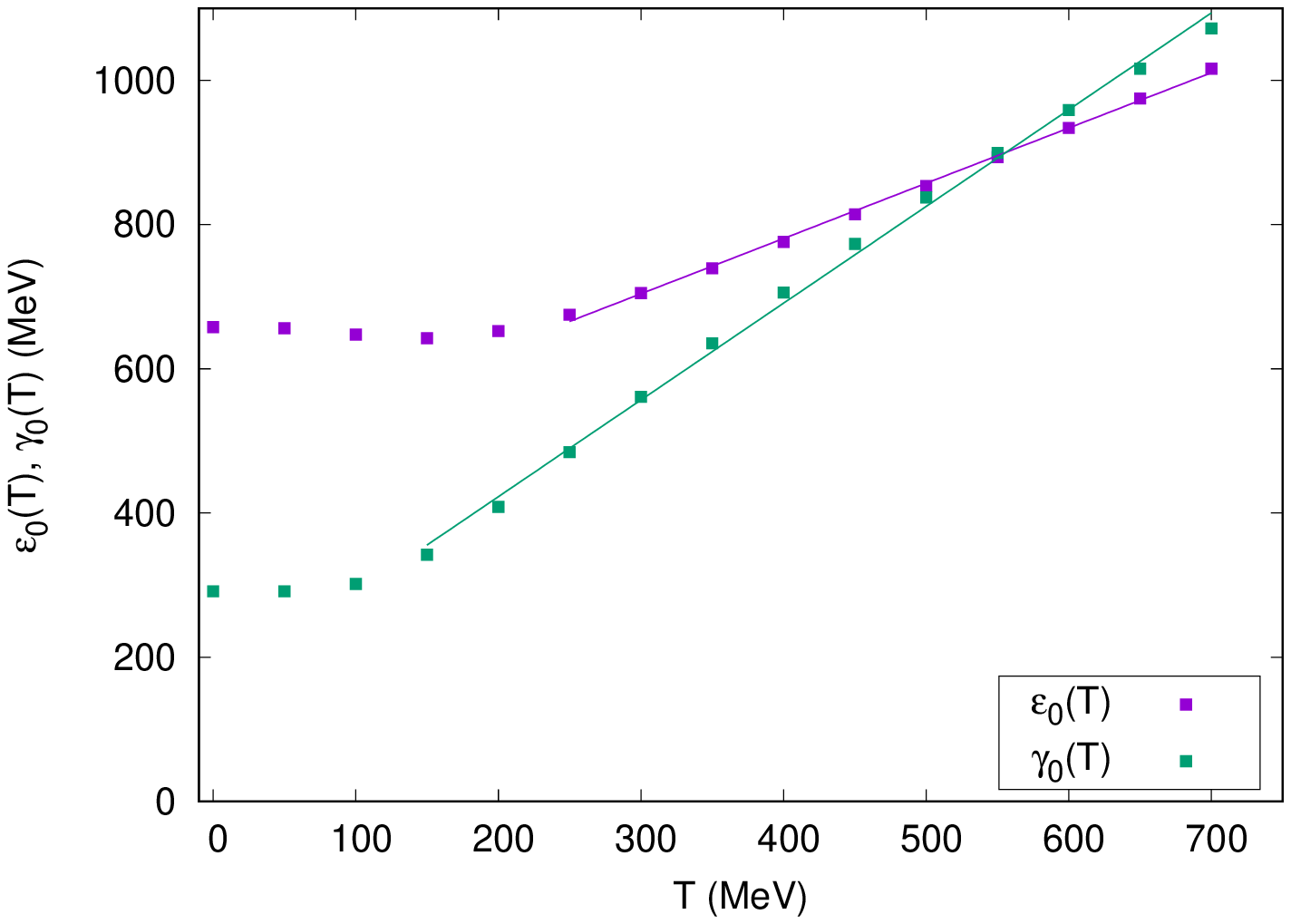}
    \caption{Real and imaginary part of one of the four symmetric, complex-conjugate poles $z(T)=i[\varepsilon_{0}(T)+i\gamma_{0}(T)]$ of the Euclidean Landau-gauge full-QCD gluon propagator at zero spatial momentum $|{\bf p}|=0$, as functions of the temperature $T$. Computed for $m=0.656$~GeV, $\pi_{0}=-0.876$, $n_{F}=2+1$, $M_{1}=0.350$~GeV, $M_{2}=0.450$~GeV.}
    \label{poles_qcd_fixed}
\end{figure}

In Fig.~\ref{poles_qcd_fixed} we display the real and the imaginary part of the gluon poles in $n_{f}=2+1$ full QCD computed using the parameters obtained by optimizing pure Yang-Mills theory in vacuum -- namely, $m=0.656$~GeV, $\pi_{0}=-0.876$, like in Sec.~IIIA. As discussed in Sec.~IIA, we model quark mass generation due to chiral symmetry breaking by setting the masses of the internal quark lines in Fig.~\ref{gluqkdiag} equal to the IR-generated mass $M_{q}\approx400$~MeV \cite{KBLW05,CRBS21}, instead of the current mass. Denoting the lightest quark pair's mass with $M_{1}$ and that of the heaviest quark with $M_{2}$, we choose $(M_{1},M_{2})=(0.350, 0.450)$~GeV to reflect a difference of about $100$~MeV between the corresponding current masses, inherited by their dressed IR limit\footnote{The actual difference in the IR may be smaller -- see e.g. \cite{CRBS21} -- but we keep it this large to distinguish the effects of the heaviest quark from that of the lightest pair.}. These values~-- and the model itself -- were already used in \cite{CS25} to study the behavior of the static gluon propagator in full QCD as a function of temperature and baryonic density.

The evolution of $\varepsilon_{0}(T)$ and $\gamma_{0}(T)$ with temperature displayed in Fig.~\ref{poles_qcd_fixed} is similar to that of its pure-Yang-Mills analogue of Sec.~IIIA~-- see Fig.~\ref{poles_ym_fixed}. The real part $\varepsilon_{0}(T)$ shows a small decrease from $\varepsilon_{0}(T=0)=658$~MeV to $\varepsilon_{0}(T)\approx642$~MeV at $T\approx 145$~MeV, followed by an increase roughly linear with temperature for $T\gtrapprox250$~MeV. The latter is well parametrized by the function
\begin{equation}\label{qcdfixedlineare}
    \varepsilon_{0}(T)\approx 474.1\text{ MeV} + 0.7664\, T\quad(T\in[250,700]\text{ MeV})\ .
\end{equation}
Over the whole temperature range, $\varepsilon_{0}(T)$ is larger than in pure Yang-Mills theory. At high temperatures, where the linear coefficient in Eq.~\eqref{qcdfixedlineare} is larger by a factor of $1.65$, this behavior is consistent with the known increase of the plasmon frequency $\omega_{P}$ of one-loop ordinary perturbation theory \cite{Haque04},
\begin{equation}
    \omega_{P}^{2}=\frac{g^{2}T^{2}}{9}\,\left(N+\frac{n_{f}}{2}\right)\ ,
\end{equation}
with the number of fermions $n_{f}$, despite the fact that, for $T\leq 700$~MeV, with a gluon mass parameter $m\sim T$ and quark masses of the same order, the approximation provided by the above equation clearly does not hold.

When computed at fixed, optimized parameters, the imaginary part of the full-QCD gluon pole $\gamma_{0}(T)$ slowly increases from $\gamma_{0}(T=0)=291$~MeV up to $T\approx100$-$150$~MeV, then it becomes approximately linear with temperature at $T\gtrapprox150$~MeV, with an intercept and a slope given by
\begin{equation}\label{qcdfixedlinearg}
    \gamma_{0}(T)\approx 154.3\text{ MeV} + 1.3417\,T\quad(T\in[150,700]\text{ MeV}).
\end{equation}
While its $T\to0$ limit is sensibly smaller than in pure Yang-Mills theory, $\gamma_{0}(T)$ eventually exceeds its $n_{f}=0$ counterpart at higher temperatures, as indicated by the linear coefficient in Eq.~\eqref{qcdfixedlinearg} being larger by a factor of 1.22 than in Eq.~\eqref{ymfixedlinearg}. The fact that the ratio between the imaginary parts is smaller than that of the real parts can also be understood in perturbative terms: at high temperatures, when computed in ordinary perturbation theory, the quark contribution to $\gamma_{0}(T)$ is known to be effectively $O(g^{3})$, while that of gluons is $O(g^{2})$ \cite{Heinz87}; this makes $\gamma_{0}(T)$ less sensitive to a change in $n_{f}$ than $\varepsilon_{0}(T)$. Again, we stress that this is only a qualitative argument: the temperatures considered in this study are of the same order of the gluon and quark masses, so ordinary massless perturbation theory can only serve us to make rough comparisons. A larger $\varepsilon_{0}(T)$, combined with smaller values of $\gamma_{0}(T)$ at low temperatures and to a larger increase in the slope of the former than in that of the latter, all contribute to the real and imaginary part of the gluon pole crossing at a higher temperature~-- namely $T\approx550$~MeV~-- than observed in fixed-parameter pure Yang-Mills theory.

\begin{figure}[h]
    \centering
    \includegraphics[width=0.33\textwidth,angle=270]{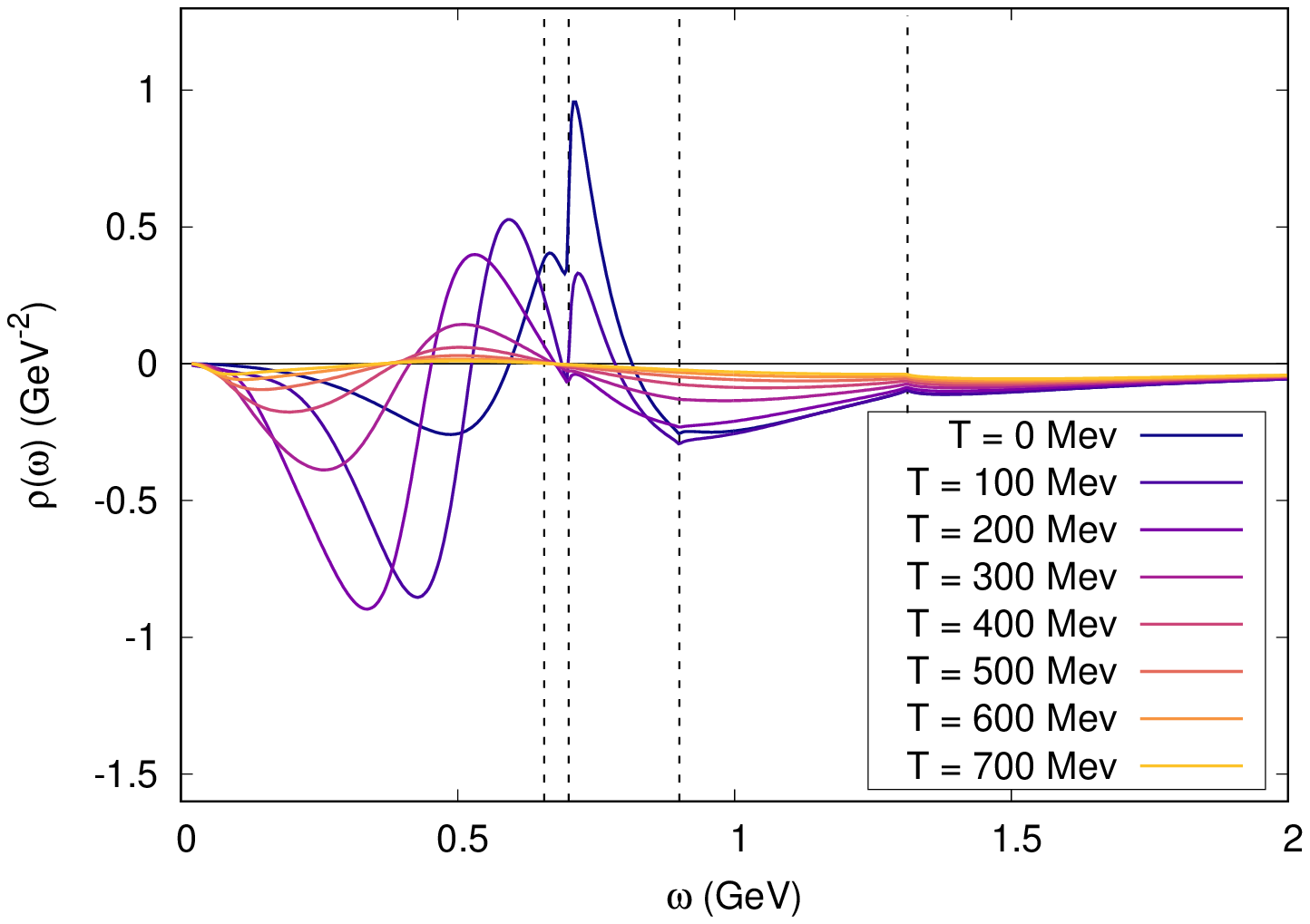}
    \caption{Zero-momentum ($|{\bf p}|=0$) spectral function of the Landau-gauge $n_{f}=2+1$ full-QCD gluon propagator as a function of the spectral frequency $\omega$, for different values of the temperature $T$. The dashed vertical lines mark the $m$, $2M_{1}$, $2M_{2}$ and $2m$ mass thresholds. Computed for $m=0.656$~GeV, $\pi_{0}=-0.876$, $M_{1}=0.350$~GeV, $M_{2}=0.450$~GeV.}
    \label{spectral_qcd_fixed}
\end{figure}

In Fig.~\ref{spectral_qcd_fixed} we display the zero-momentum full-QCD $n_{f}=2+1$ spectral function $\rho(\omega)=\rho_{T,L}(\omega,|{\bf p}|=0)$ computed using the pure-Yang-Mills optimized vacuum parameters. As is clear by a comparison with Figs.~\ref{spectral_ym_fixed} and \ref{spectral_ym_lattice}, $\rho(\omega)$ in full QCD has a much richer structure than in pure Yang-Mills theory. This is a direct consequence of the additional singularities brought about by the quark loops in Fig.~\ref{gluqkdiag}: each of these contains two internal quark lines of mass $M_{i}$ ($i=1,2$) -- that is, two massive denominators of the form $1/(k^{2}+M^{2}_{i})$ --, which translates to new contributions to the propagator's branch cut at the mass thresholds $p^{2}<-4M_{1}^{2}$, $p^{2}<-4M_{2}^{2}$. In terms of the zero-momentum spectral function, these manifest themselves as singularities at $|\omega|=2M_{1}$ and $|\omega|=2M_{2}$. Thus, for our choice of parameters, the full-QCD $|{\bf p}|=0$ spectral function is singular at $\omega=m=0.656$~GeV, $\omega=2M_{1}=0.700$~GeV, $\omega=2M_{2}=0.900$~GeV and $\omega=2m=1.312$~GeV. These four thresholds are marked with dashed vertical lines in the figure. Just as in pure-Yang-Mills theory, terms which are non-analytic at $\omega=m$ are multiplied by powers of $p^{2}+m^{2}$, so corresponding discontinuities only affect higher-order derivatives.

For $\omega\lessapprox m$ and $\omega> 2m$, the behavior of the spectral function is similar to pure Yang-Mills theory at optimized $T=0$ parameters. $\rho(\omega)$ is negative at large $\omega$, increasing towards zero with temperature, and shows a low-$\omega$ minimum followed by a maximum whose depth, height and positions depend on the temperature the way we described in Sec.~IIIA: their position shifts down to lower $\omega$ as $T$ increases, while their values are non-monotonic with temperature, the minimum being deepest at $T\approx150$~MeV and the maximum highest at $T\approx130$~MeV.

For $\omega>m$, an $\omega\approx 2M_{1}$ local maximum lies above the corresponding threshold\footnote{Although not clear from the figure, it can be verified numerically that the maximum persists up to the highest considered temperature $T=700$~MeV due to the derivative $\partial\rho/\partial\omega$ being positive as $\omega\to 2M_{1}^{+}$.}. For sufficiently low $T$, the range of frequencies where the spectral function is positive encompasses the two $\omega\sim m$ and $\omega\approx 2M_{1}$ maxima. As the temperature increases, the $m\approx2M_{1}$ maximum first quickly drops below the $\omega\sim m$ maximum -- the two of them being equal at $T\approx 85$~MeV --, then becomes negative at $T\approx 160$~MeV, and finally it grows back towards zero without ever becoming positive again over the considered temperature range. At the same time, the positivity interval shifts to lower frequencies like in fixed-parameter pure Yang-Mills theory. Between $\omega=2M_{2}$ and $\omega=2m$, the spectral function is negative and monotonically increasing with temperature towards the upper end of the interval, whereas around the $\omega=2M_{2}$ threshold it shows a non-monotonic behavior: it decreases with $T$ below $T\approx120$~MeV, and increases to zero at higher temperatures.

\subsection{Temperature-dependent lattice parameters}

In this section we display the poles and spectral function of the Landau-gauge gluon propagator computed in $n_{f}=2$ full QCD using the parameters in Tab.~\ref{qcd_lattice_params}. These were obtained by fitting a recently reported set of unquenched lattice data for the static limit ($\omega_{n}=0$) of the transverse component of the gluon propagator in $n_{f}=2$ full QCD \cite{SOS25}. The latter use a non-perturbative $\mc{O}(a)$-improved Wilson action for quarks, with bare lattice quark masses $m_{q,\text{bare,latt.}}=8$~MeV, yielding a pion of mass $m_{\pi}=290$~MeV. More details on the lattice setup can be found in \cite{OSSS19}. As reported in the introduction to Sec.~IV, the fit was found not to be very sensitive on the quark mass $M_{q}$. In what follows we will display results obtained by fixing $M_{q}=400$~MeV across all temperatures.

\begin{table}
    \setlength{\tabcolsep}{9pt}
    \begin{tabular}{|c|c|c|}
        \hline
        $T$ (MeV) & $m(T)$ (MeV) & $\pi_{0}(T)$ \\
        \hline
        139 & 751.8 & -0.6932\\
        154 & 763.9 & -0.6325\\
        174 & 734.5 & -0.5644\\
        199 & 730.6 & -0.5601\\
        233 & 746.3 & -0.5434\\
        278 & 792.5 & -0.5294\\
        \hline
    \end{tabular}
    \caption{Values of the parameters $m(T)$ and $\pi_{0}(T)$ obtained by fitting the lattice data of \cite{SOS25} for the transverse component of the Landau-gauge Euclidean $n_{f}=2$ full-QCD gluon propagator at zero Matsubara frequency ($\omega_{n}=0$), renormalized at $\mu=4$~GeV in the MOM scheme (see Sec.~IIB), using a quark mass $M_{q}=400$~MeV.}
    \label{qcd_lattice_params}
\end{table}

\begin{figure}[h]
    \centering
    \includegraphics[width=0.33\textwidth,angle=270]{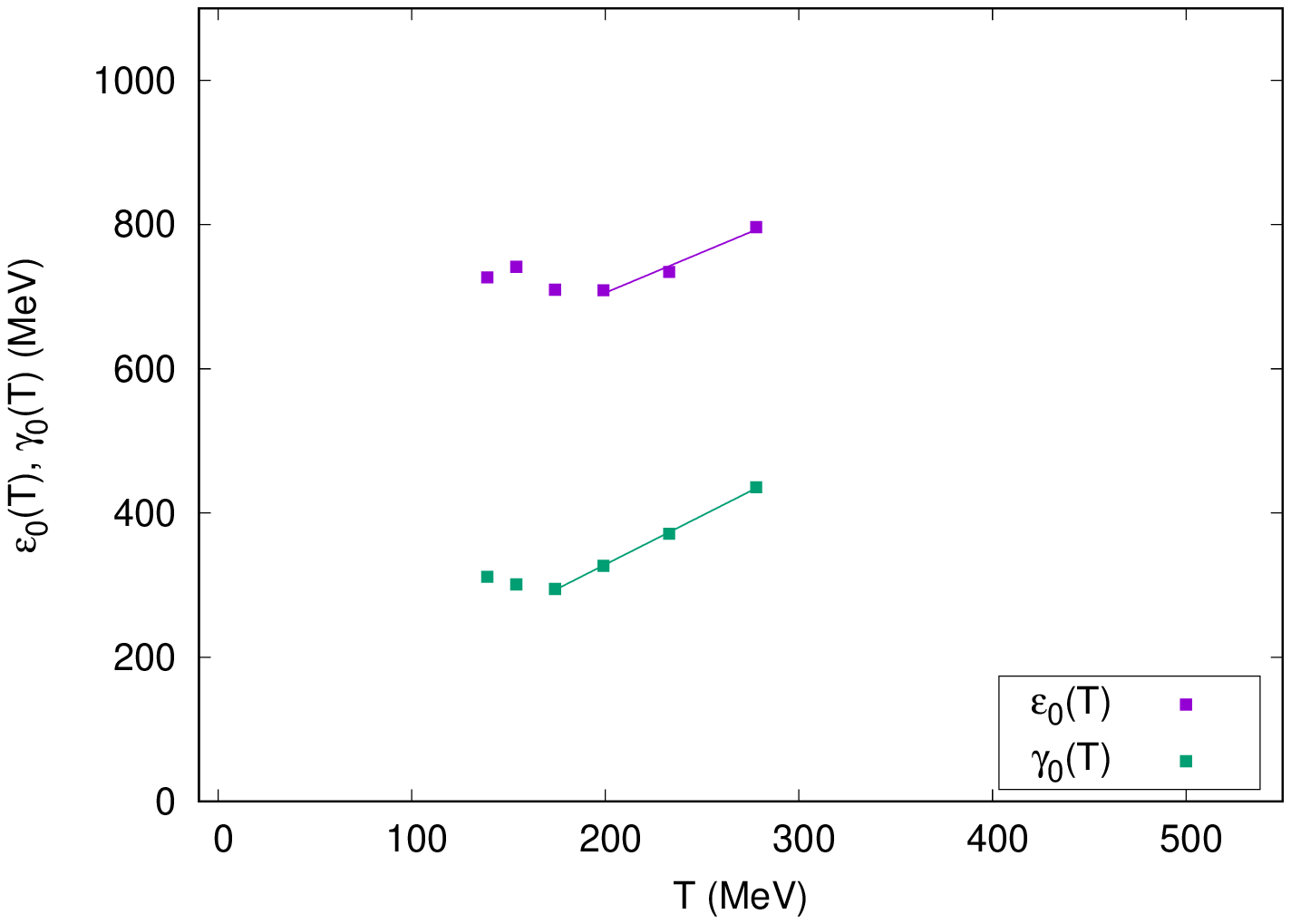}
    \caption{Real and imaginary part of one of the four symmetric, complex-conjugate poles $z(T)=i[\varepsilon_{0}(T)+i\gamma_{0}(T)]$ of the Euclidean Landau-gauge $n_{f}=2$ full-QCD gluon propagator at zero spatial momentum $|{\bf p}|=0$, as functions of the temperature $T$. Computed using the parameters in Tab.~\ref{qcd_lattice_params}.}
    \label{poles_qcd_lattice}
\end{figure}

Fig.~\ref{poles_qcd_lattice} shows the zero-momentum gluon poles $z(T)=i[\varepsilon_{0}(T)+i\gamma_{0}(T)]$ as a function of temperature. Regardless of the decrease of $n_{f}$ from 3 to 2, the increased value of the gluon mass parameter $m$ results in a net increase in the value of the real part $\varepsilon_{0}(T)$ with respect to its fixed-parameter analogue of Sec.~IVA: $\varepsilon_{0}(T)=(727,796)$~MeV at the boundaries $T=(139,278)$~MeV of the considered temperature range, against $\varepsilon_{0}(T)=(642,691)$~MeV when computed at the same temperature using optimized parameters. The increase is more marked when compared with the pure-Yang-Mills results of Sec.~IIIB -- e.g. $\varepsilon_{0}(T)=(590,674)$~MeV at $T=(121,290)$~MeV --, for which both $n_{f}$ (trivially) and $m$ (at comparable temperatures) are smaller. The imaginary part $\gamma_{0}(T)$ shows a tendency to decrease with temperature towards $T=T_{c}\approx158$~MeV, the critical temperature of the deconfinement crossover reported in \cite{SOS25} for the present lattice setup. With our fitted parameters, however, its minimum is actually attained at a somewhat larger temperature, $T= 174$~MeV, where $\gamma_{0}(T)=295$~MeV. Over the considered temperature range, $\gamma_{0}(T)$ is smaller than both fixed-parameter $n_{f}=2+1$ QCD (Sec.~IVA) and pure Yang-Mills theory using parameters fitted from the lattice (Sec.~IIIB): $\gamma_{0}(T)=(312,436)$~MeV at the boundaries $T=(139,278)$~MeV, against $\gamma_{0}(T)=(330,528)$~MeV in optimized-parameter $n_{f}=2+1$ full QCD at the same temperatures and $\gamma_{0}(T)=(383,450)$~MeV at $T=(121,290)$~MeV in pure Yang-Mills theory with lattice parameters. 

$\varepsilon_{0}(T)$ and $\gamma_{0}(T)$ in Fig.~\ref{poles_qcd_lattice} show a linear behavior with temperature for $T>T_{c}$ -- with that of the former setting in at $T\gtrapprox199$~MeV and that of the latter at $T\gtrapprox174$~MeV~-- which can be parametrized by
\begin{equation}\label{qcd_lat_linearfit_e}
    \varepsilon_{0}(T)\approx481.2\text{ MeV} + 1.1209\,T\quad(T\in[199,278]\text{ MeV})\ ,
\end{equation}
\begin{equation}\label{qcd_lat_linearfit_g}
    \gamma_{0}(T)\approx 57.1\text{ MeV} + 1.3572\,T\quad(T\in[174,278]\text{ MeV})\ .
\end{equation}
Like in Sec.~IIIB, we must remark that these linear regressions were obtained by making use of a limited number of data points -- three and four for $\varepsilon_{0}(T)$ and $\gamma_{0}(T)$, respectively. Additionally, here, the temperature range involved in the regression is narrower than in pure Yang-Mills theory. Therefore these results should be used with care when trying to extrapolate to higher temperatures. For instance, the fact that the fitted intercept of $\gamma_{0}(T)$ is larger than that of $\varepsilon_{0}(T)$ inevitably causes the former to exceed the latter at a sufficiently high temperature. A trivial calculation shows this happens at $T=1795$~MeV, which is well beyond the domain of the analysis carried out in the present study. We can thus conclude that, when the gluon propagator is computed to one loop in the framework of the screened massive expansion of full QCD using temperature-dependent parameters obtained from the lattice data of \cite{SOS25}, the real and the imaginary part of its poles are found not to cross within the domain of the present analysis, at variance with the corresponding calculation carried out at temperature-independent optimized parameters (Sec.~IVA) and in agreement with what we already observed in Sec.~IIIB in the context of pure Yang-Mills theory using the lattice data of \cite{SOBC14}.

\begin{figure}[h]
    \centering
    \includegraphics[width=0.33\textwidth,angle=270]{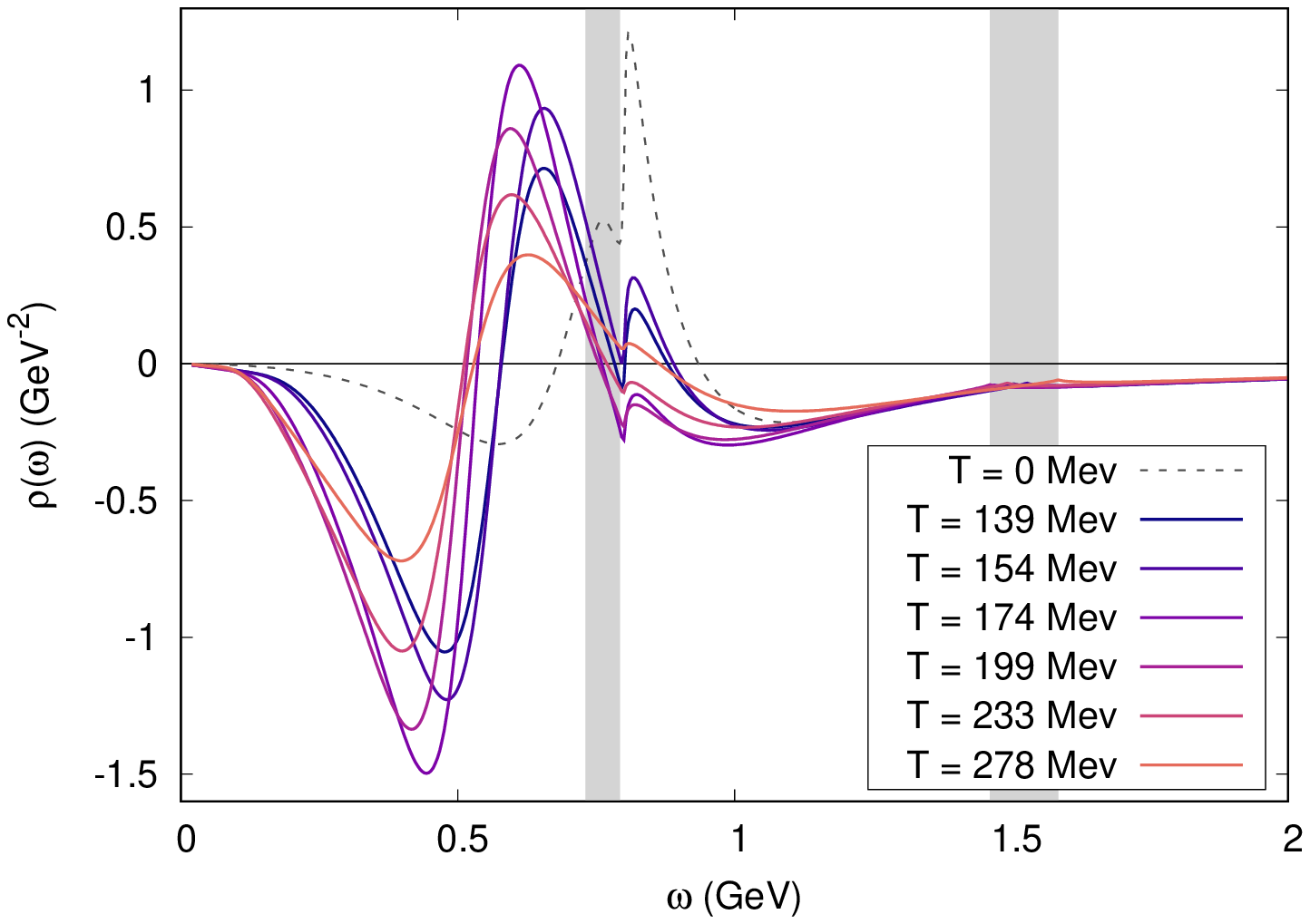}
    \caption{Zero-momentum ($|{\bf p}|=0$) spectral function of the Landau-gauge $n_{f}=2$ full-QCD gluon propagator as a function of the spectral frequency $\omega$, for different values of the temperature $T$. Computed using the parameters in Tab.~\ref{qcd_lattice_params}. The grey bands mark the regions $\omega\in[m_{\text{min}},m_{\text{max}}]$ and $\omega\in[2m_{\text{min}},2m_{\text{max}}]$, where $m_{\text{min,max}}$ are the minimum and maximum gluon mass parameter in the table. The dashed curve, included for comparison, was evaluated at $T=0$ using the lowest-temperature ($T=139$~MeV) parameters from the lattice.}
    \label{spectral_qcd_lattice}
\end{figure}

Fig.~\ref{spectral_qcd_lattice} shows the $n_{f}=2$ full-QCD zero-momentum Landau-gauge spectral function $\rho(\omega)=\rho_{T,L}(\omega,|{\bf p}|=0)$ computed using the lattice parameters in Tab.~\ref{qcd_lattice_params}. The dashed black line is added for the purpose of comparison and displays the spectral function evaluated at $T=0$ using the same parameters as the lowest-temperature curve, $T=139$~MeV. We remark that, at variance with the $T=0$ curve in Fig.~\ref{spectral_ym_lattice}, these parameters were not obtained by fitting lattice data for the displayed temperature, which motivates our different choice of color and line coding.

In the presence of a pair of quarks of equal mass $M_{q}=400$~MeV, the spectral function possesses three singular points at $\omega=m$, $\omega=2M_{q}=800$~MeV and $\omega=2m$, corresponding to analogous mass thresholds. As in the previous section, the $\omega=m$ singularity is a discontinuity in higher-order derivatives of $\rho(\omega)$, whereas those at $\omega=2M_{q}$ and $\omega=2m$ are discontinuities in $\partial\rho/\partial\omega$. Due to our choice of parameters, the $\omega=2M_{q}$ threshold is temperature independent, while the singularities at $\omega=m,2m$ depend on $T$ as indicated in Tab.~\ref{qcd_lattice_params}.

Despite the different analytic structure -- reflecting the different quark content~--, the spectral function in Fig.~\ref{spectral_qcd_lattice} shares a number of features with that of Sec.~IVA. First of all, it displays similar local extrema: it has a minimum for $\omega\lessapprox0.5$~GeV, followed by a maximum at $\omega\sim m$, by a second maximum right above the $\omega=2M_{q}$ threshold and by another minimum for $\omega<2m$. Moreover, it is negative at low and high $\omega$'s, and its positivity interval encompasses the $\omega\sim m$ and~-- at low temperatures -- the $\omega\sim 2M_{q}$ maxima. On the other hand, the $\omega\approx 2M_{q}$ maxima do not show the clear -- albeit non-monotonous -- ordering with temperature displayed by the fixed-parameter $n_{f}=2+1$ calculation of Sec.~IVA: the value of $\rho(\omega)$ at the maximum first increases with temperature between $T=139$~MeV and $T=154$~MeV, then decreases up to a temperature between $199$~MeV and $233$~MeV, then it increases again. At $T=278$~MeV it becomes positive, which is a feature not displayed by the fixed-parameter calculations even at much higher temperatures. This behavior can be attributed to the interplay between temperature and mass effects: at fixed quark mass and temperature, the value of the $\omega\approx2M_{q}$ maximum is found to increase with the gluon mass $m$; this explains both its increase between $T=139$~MeV and $T=154$~MeV -- where mass effects dominate over the decrease with temperature observed in Sec.~IVA -- and its becoming positive at $T=278$~MeV~-- where mass effects amplify the increase with temperature already in action at larger $T$'s. Finally, we observe that, just like in Sec.~IIIB, both the $\omega\lesssim m$ minima and the $\omega \sim m$ maxima show a tendency to line up as a result of the overall increase of the gluon mass parameter $m$ with temperature, which compensates the shift towards lower $\omega$'s seen at fixed $m$ in Sec.~IVA\footnote{It is easy to see from Fig.~\ref{spectral_qcd_lattice} and Tab.~\ref{qcd_lattice_params} that the position of the $\omega\lesssim m$ minima and that of the $\omega \sim m$ maxima only decreases noticeably when $m$ decreases.}: the former is located between $\omega=0.39$~GeV and $\omega=0.48$~GeV, whereas the latter lies between $\omega=0.59$~GeV and $\omega=0.66$~GeV; in absolute value, they are both larger in the interval $T\in [154,199]$~MeV.

\subsection{Other quark masses}

In light of the anticipated dependence of the quark mass scale on temperature, our analysis cannot exempt itself from exploring the possibility that a different analytic structure for the gluon propagator may arise from a change in such a scale. Of particular importance in this context is understanding how the gluon propagator responds to a decrease in the quark mass, which is expected to take place in the deconfined phase right above the crossover temperature $T=T_{c}\approx 150$~MeV because of chiral symmetry restoration. Starting from the vacuum infrared scale $M_{q}\approx400$~MeV, a decrease in $M_{q}$ would soon push the $2M_{q}$ two-particle threshold below the gluon mass scale $m\approx700$~MeV, which itself is predicted to increase with temperature beyond $T_{c}$ both by ordinary perturbative arguments and by our fits, see Tab.~\ref{qcd_lattice_params}. From the results presented in the previous sections, one can already anticipate that the spectral function of the one-loop full-QCD gluon propagator will be sensitive to this change. At higher temperatures, on the other hand, developing an intuition on the ordering of mass scales is far harder due to the fact that both $M_{q}$ and $m$ are expected to increase with $T$: it is unclear then whether $M_{q}$ will again be such that $m<2M_{q}<2m$ as in Secs.~IVA-B, or if the thresholds will exchange places.

In the absence of a quantitative model for the evolution of mass scales at high temperatures, we will not try to pursue a detailed description of the poles and spectral function of the propagator at larger $M_{q}$'s. Nonetheless, we did verify using the optimized parametrization that, up to temperatures $T=700$~MeV, the overall analytic structure of the one-loop full-QCD gluon propagator computed in the screen massive expansion does not depend on the mass of the lightest quarks over a wide interval, spanning from $M_{q}=2$~MeV to $M_{q}=2$~GeV: regardless of $M_{q}$, the propagator is still found to have two pairs of symmetric complex-conjugate poles in the complexified frequency $\omega\to z\in\mathbb{C}$ and the ordinary branch cut along the imaginary (Minkowski) axis; furthermore, the spectral function was not found to develop sharp peaks. As $M_{q}$ increases and becomes larger than $m$, the $2M_{q}$ singularities shift to higher frequencies and step over the $2m$ threshold; at large enough $M_{q}$'s, in the deep infrared, both the spectral function and the poles approach their pure-Yang-Mills limit. In what follows, we will limit ourselves to displaying results on the poles and spectral function of the gluon propagator computed for $2M_{q}<m$, where the different ordering of mass thresholds and smallness of $M_{q}$ translate to a different behavior than described in Secs.~IVA-B for full QCD and in Secs.~IIIA-B for pure Yang-Mills theory. For this purpose, we will choose $(M_{1},M_{2})=(125,225)~$MeV like we did in \cite{CS25} for an alternative fixed, optimized $n_{f}=2+1$ calculation, and set $M_{q}=200$~MeV to perform a new fit of the $n_{f}=2$ lattice data of \cite{SOS25}.\\

\begin{table}
    \setlength{\tabcolsep}{9pt}
    \begin{tabular}{|c|c|c|}
        \hline
        $T$ (MeV) & $m(T)$ (MeV) & $\pi_{0}(T)$ \\
        \hline
        139 & 736.0 & -0.7693\\
        154 & 749.3 & -0.7069\\
        174 & 721.7 & -0.6386\\
        199 & 719.7 & -0.6291\\
        233 & 737.6 & -0.6044\\
        278 & 785.6 & -0.5790\\
        \hline
    \end{tabular}
    \caption{Values of the parameters $m(T)$ and $\pi_{0}(T)$ obtained by fitting the lattice data of \cite{SOS25} for the transverse component of the Landau-gauge Euclidean $n_{f}=2$ full-QCD gluon propagator at zero Matsubara frequency ($\omega_{n}=0$), renormalized at $\mu=4$~GeV in the MOM scheme (see Sec.~IIB), using a quark mass $M_{q}=200$~MeV.}
    \label{qcd_lattice_params_lowmass}
\end{table}

\begin{figure}[h]
    \centering
    \includegraphics[width=0.33\textwidth,angle=270]{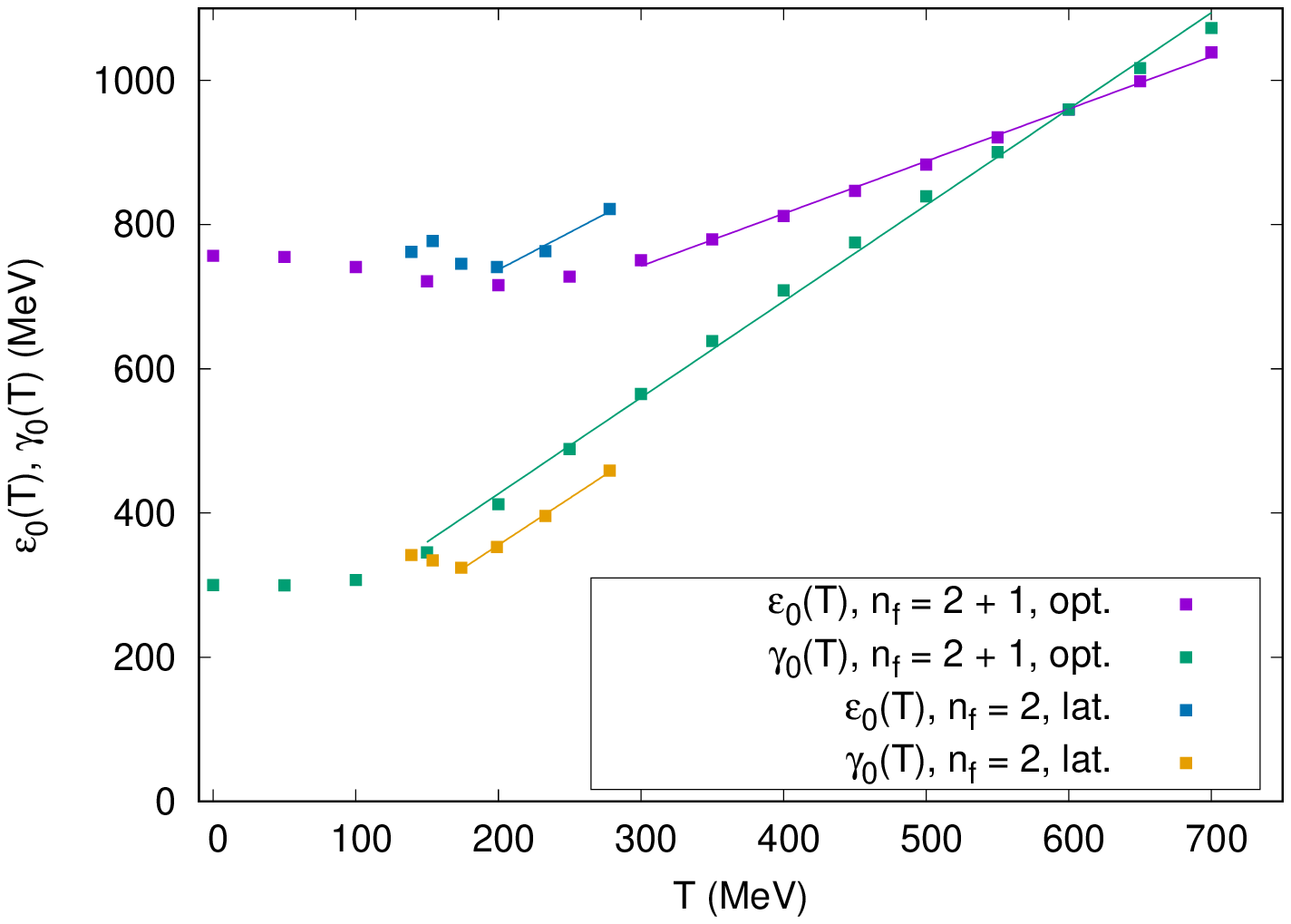}
    \caption{Real and imaginary part of one of the four symmetric, complex-conjugate poles $z(T)=i[\varepsilon_{0}(T)+i\gamma_{0}(T)]$ of the Euclidean Landau-gauge full-QCD gluon propagator at zero spatial momentum $|{\bf p}|=0$, as functions of the temperature $T$. See the main text for details.}
    \label{poles_qcd_lowmass}
\end{figure}

In Fig.~\ref{poles_qcd_lowmass} we show the real and imaginary part of the zero-momentum Landau-gauge gluon poles computed both in $n_{f}=2+1$ full QCD at fixed Yang-Mills optimized parameters, and in $n_{f}=2$ full QCD using the temperature-dependent parameters reported in Tab.~\ref{qcd_lattice_params_lowmass}. As mentioned above, the latter were obtained by fitting the screened massive expansion's propagator to the lattice data of \cite{SOS25} using a quark mass $M_{q}=200$~MeV, whereas for the $n_{f}=2+1$ optimized calculation we used quark masses $(M_{1},M_{2})=(125,225)$~MeV. In both cases, $2M_{i}<m$ ($i=1,2,q$).

As is clear by a comparison with Figs.~\ref{poles_qcd_fixed} and \ref{poles_qcd_lattice}, lowering the quark masses to about half their initial value does not modify the qualitative behavior of the gluon poles with temperature. Quantitatively, the most relevant difference is an increase in the real part $\varepsilon_{0}(T)$ of the optimized-parameter calculation in the low-temperature regime: at $T=0$, where $\varepsilon_{0}(T)=658$~MeV when computed with larger quark masses, the real part is $\varepsilon_{0}(T)=757$~MeV. On the other hand, at higher temperatures, where a linear regression of the optimized-parameter poles yields
\begin{equation}
    \varepsilon_{0}(T) \approx 524.6 + 0.7263\,T\quad(T\in[300,700]\text{ MeV})\ ,
\end{equation}
\begin{equation}
    \gamma_{0}(T) \approx 159.6 + 1.3346\,T\quad(T\in[150,700]\text{ MeV})\ ,
\end{equation}
the difference is more contained, decreasing to as little as $\approx 22$~MeV at $T=700$~MeV. The same difference evaluated the for poles computed at temperature-dependent lattice parameters is of a similar order: in going from $M_{q}=400$~MeV (Tab.~\ref{qcd_lattice_params}) to $M_{q}=200$~MeV (Tab.~\ref{qcd_lattice_params_lowmass}), $\varepsilon_{0}(T)$ increases by $\approx35$~MeV at $T=139$~MeV and by $\approx25$~MeV at $T=278$~MeV. A comparable increase is found for the imaginary part $\gamma_{0}(T)$ computed at lattice parameters, which differs from our previous determination by $\approx30$~MeV and $\approx23$~MeV at the lower and upper ends of the temperature interval, whereas the same difference computed at optimized parameters is practically negligible, going from $\approx9$~MeV at $T=0$ to less than $1$~MeV at $T=700$~MeV. For completeness, we also report the high-temperature linear regressions of the real and imaginary part of the gluon poles computed using the parameters in Tab.~\ref{qcd_lattice_params_lowmass}:
\begin{equation}
    \varepsilon_{0}(T) \approx 530.7 + 1.0332\,T\quad(T\in[199,278]\text{ MeV})\ ,
\end{equation}
\begin{equation}
    \gamma_{0}(T) \approx 95.8 + 1.2985\,T\quad(T\in[174,278]\text{ MeV})\ .
\end{equation}

\begin{figure}[h]
    \centering
    \includegraphics[width=0.33\textwidth,angle=270]{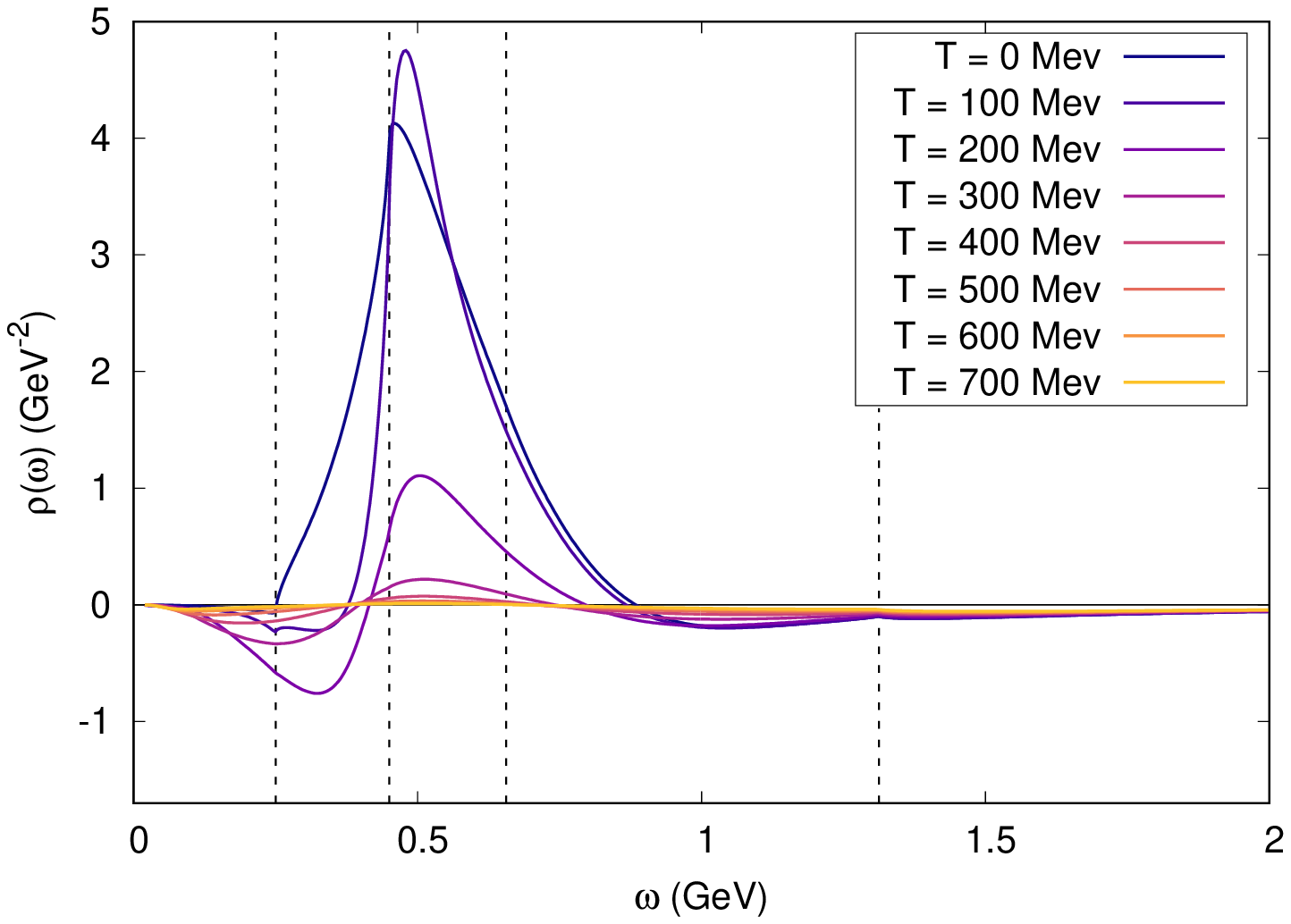}
    \caption{Zero-momentum ($|{\bf p}|=0$) spectral function of the Landau-gauge $n_{f}=2+1$ full-QCD gluon propagator as a function of the spectral frequency $\omega$, for different values of the temperature $T$. The dashed vertical lines mark the $2M_{1}$, $2M_{2}$, $m$ and $2m$ mass thresholds. Computed for $m=0.656$~GeV, $\pi_{0}=-0.876$, $M_{1}=0.125$~GeV, $M_{2}=0.225$~GeV.}
    \label{spectral_qcd_fixed_lowmass}
\end{figure}

\begin{figure}[h]
    \centering
    \includegraphics[width=0.33\textwidth,angle=270]{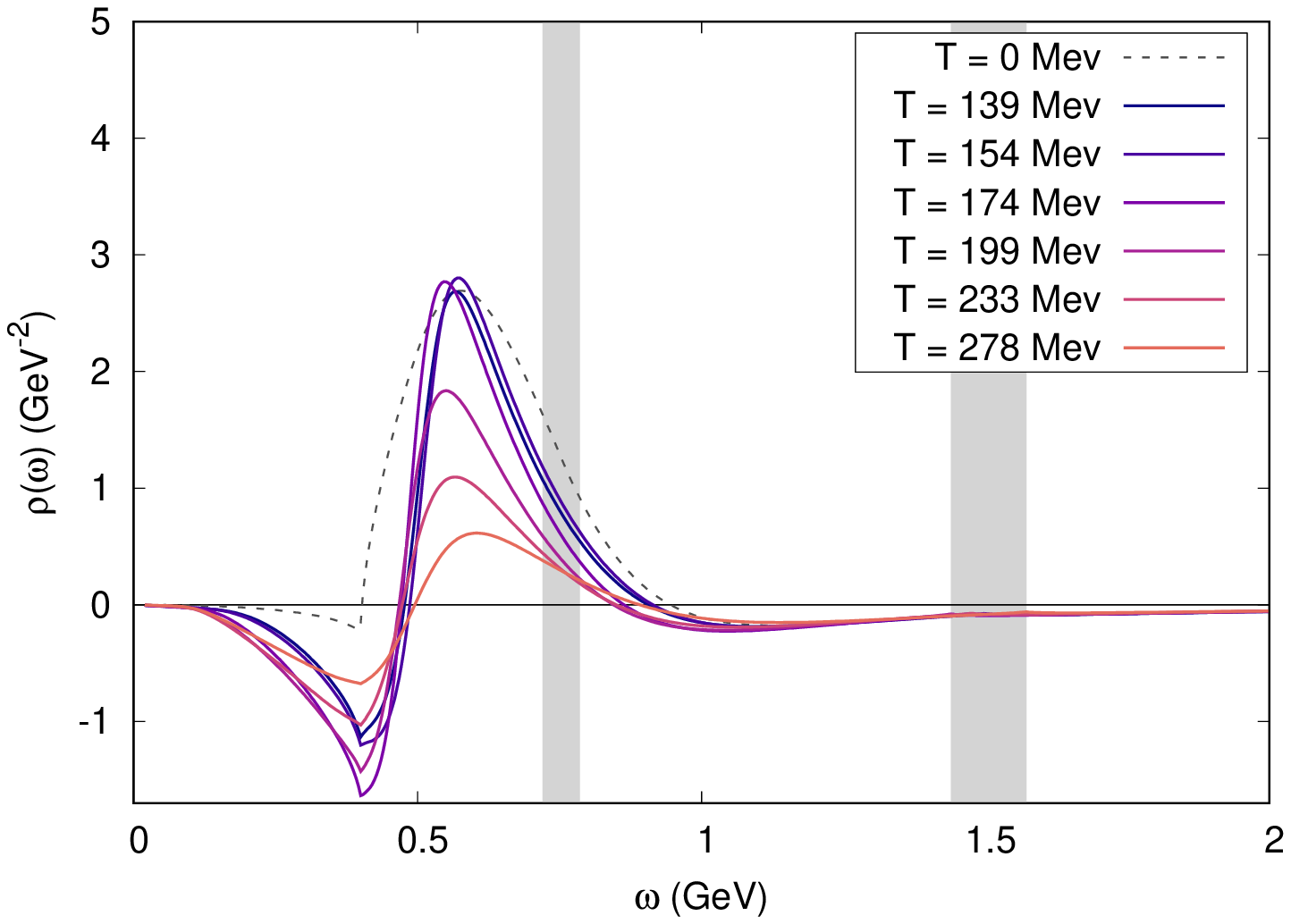}
    \caption{Zero-momentum ($|{\bf p}|=0$) spectral function of the Landau-gauge $n_{f}=2$ full-QCD gluon propagator as a function of the spectral frequency $\omega$, for different values of the temperature $T$. Computed using the parameters in Tab.~\ref{qcd_lattice_params_lowmass}. The grey bands mark the regions $\omega\in[m_{\text{min}},m_{\text{max}}]$ and $\omega\in[2m_{\text{min}},2m_{\text{max}}]$, where $m_{\text{min,max}}$ are the minimum and maximum gluon mass parameter in the table. The dashed curve, included for comparison, was evaluated at $T=0$ using the lowest-temperature ($T=139$~MeV) parameters from the lattice.}
    \label{spectral_qcd_lattice_lowmass}
\end{figure}

In Fig.~\ref{spectral_qcd_fixed_lowmass} we display the $n_{f}=2+1$ zero-momentum spectral function $\rho(\omega)=\rho_{T,L}(\omega,|{\bf p}|=0)$ computed at fixed, optimized parameters using $(M_{1},M_{2})=(125,225)$~MeV. Like before, the dashed vertical lines mark the $2M_{1}=0.250$~GeV, $2M_{2}=0.450$~GeV, $m=0.656$~GeV and $2m=1.312$~GeV mass thresholds, which however are now ordered differently than in Sec.~IVA. As expected from the shift of the first two below $\omega=m$, the low-temperature behavior of $\rho(\omega)$ with the spectral frequency $\omega$ is dissimilar to that analyzed in the previous sections. The two-maxima structure that resulted from a contribution of the lightest quarks turning on at $\omega=2M_{1}$ right above the $\omega\approx m$ maximum is replaced by a sudden increase in $\rho(\omega)$\footnote{Note the different scale between Fig.~\ref{spectral_qcd_fixed_lowmass} and the previous figures.}. The latter is primarily driven by said contribution activating at lower spectral frequencies than before and adding to the pre-existing maximum\footnote{For comparison, see Fig.~\ref{spectral_qcd_lattice_lowmass} ahead, in which the third heavier quark is not present.}, and is enhanced by effects due to the $2M_{2}$ threshold being superimposed to said maximum. The increase in $\rho(\omega)$ also prevents the smooth lowest-frequency minimum observed in both Yang-Mills theory and full QCD at higher quark masses from appearing at temperatures below $T\approx100$~MeV. The minimum is instead replaced by the discontinuity in $\partial\rho/\partial\omega$, until it starts developing again between $\omega=2M_{1}$ and $\omega=2M_{2}$ at $T\gtrapprox100$~MeV. At higher temperatures $T\gtrapprox300$~MeV, quark threshold effects -- while still present and measurable -- become less important, and the spectral function does not show a meaningfully different behavior than in pure Yang-Mills theory (Sec.~IIIA) or full QCD at higher quark masses (Sec.~IVA).

In Fig.~\ref{spectral_qcd_lattice_lowmass} we display the $n_{f}=2$ zero-momentum spectral function $\rho(\omega)=\rho_{T,L}(\omega,|{\bf p}|=0)$ computed at temperature-dependent parameters fitted from the lattice data of \cite{SOS25} using $M_{q}=200$~MeV -- see Tab.~\ref{qcd_lattice_params_lowmass}. Like in Sec.~IVB, the dashed black curve -- computed at $T=0$ using the parameters of the lowest-temperature ($T=139$~MeV) fit -- is added for reference.

In the absence of a third heavier quark, the nature of the increase in $\rho(\omega)$ above the $2M_{q}$ threshold and the subsequent lack of the two-maxima structure preceded by a smooth minimum can be understood more clearly than in Fig.~\ref{spectral_qcd_fixed_lowmass} as mainly due to the $\omega>2M_{q}$ contributions of the lightest -- and, in this case, only -- quarks to the spectral function. For the present values of $m(T)$ and $M_{q}$, the threshold itself is located in a portion of the curve $\omega < m$ where, as in our previous determinations, $\rho(\omega)$ would otherwise decrease with $\omega$; here the discontinuity in $\partial\rho/\partial\omega$ replaces the low-frequency negative minimum. At variance with Fig.~\ref{spectral_qcd_fixed_lowmass}, the spectral function does not develop such a smooth minimum for any value of temperature in the considered range.

Aside from the trivial shift of the quark mass spectral thresholds to lower $\omega$'s, the most notable features of the results presented in this section are the marked increase~-- which, however, still doesn't qualify as a sharp peak~-- of the spectral function towards positive values at frequencies $\omega\in[2M_{q},m]$, the disappearance of the smooth low-frequency minimum of $\rho(\omega)$ in favor of the singularity brought by the lighest quark threshold -- which affects some but not all choices of $(m(T), M_{q}(T))$ at a given $T$~-- and an appreciable enhancement in the real part $\varepsilon_{0}(T)$ of the zero-momentum gluon poles when all other parameters are kept fixed. All of these may be relevant over an intermediate range of temperatures located right above $T_{c}$, where the quark mass scale is expected to drop before starting to grow back again due to ordinary thermal effects.

\section{Discussion}

With the aim of looking for signatures of the deconfinement phase transition, in the present paper we have explored the deep-infrared analytic structure of the Landau-gauge gluon propagator at finite temperature, as determined by a one-loop calculation within the framework of the screened massive expansion of pure Yang-Mills theory and of QCD. By employing both parameters obtained by optimizing the pure Yang-Mills theory propagator in vacuum by principles of gauge invariance \cite{SC18}, and parameters obtained by fitting Euclidean lattice data for its static limit \cite{CS25,Com25b}, we have computed the gluon poles and spectral function at vanishing spatial momentum $|{\bf p}|=0$ over a temperature range extending from vacuum ($T=0$) to temperatures relevant to quark-gluon plasma phenomenology: $T\in[0, 700]$~MeV with optimized parameters, $T\in[121,458]$~MeV with lattice parameters in pure Yang-Mills theory, and $T\in[139,278]$~MeV with lattice parameters in full QCD. Due to the nature of the one-loop calculation and to the lack of a first-principles determination of the quark mass generated in the infrared by chiral symmetry violation, in full QCD we used the simplest possible model for the quark masses: we set them to constants of the order of the QCD scale $\Lambda_{\text{QCD}}$, choosing $M_{q}\sim400$~MeV, the mass of light quarks in the deep infrared as determined by lattice studies \cite{KBLW05}. In the framework of the screened massive expansion, this procedure can be interpreted as a truncation to the lowest-order approximation of a chirally broken vacuum, which should then be corrected by higher perturbative orders including vertices of the form $M_{q}-m_{q}$, where $m_{q}$ is the quark's current mass \cite{CRBS21}. Since the focus of the present paper is the gluon sector, we did not take these corrections into account.

As in previous studies \cite{SIR17d,SC21}, over the examined temperature range, in both pure Yang-Mills theory and full QCD, and at zero momentum $|{\bf p}|=0$, the gluon propagator evaluated to one loop in the framework of the screened massive expansion was found to have a pair of complex-conjugate poles in its squared complex frequency variable~-- equivalently, a quartet of symmetric complex-conjugate poles in frequency -- located on its principal Riemann sheet. Let us focus on the pole which, adopting the Minkowski point of view, has positive real and imaginary parts, $\varepsilon_{0}(T)$ and $\gamma_{0}(T)$, respectively. When computed using temperature-independent optimized parameters, the real part of the pole is found to slightly decrease as $T$ approaches an intermediate temperature, $T\approx165$~MeV for pure Yang-Mills theory and $T\approx 145$~MeV for full QCD; the corresponding calculations carried out with temperature-dependent lattice parameters show no such behavior within the indeterminacies of the fit to the lattice data. The imaginary part of the pole strictly increases with temperature in all but the full-QCD calculation with lattice parameters, where a slight decrease is observed up to a temperature close to $T_{c}$~-- namely, $T=174$~MeV. In all cases, at sufficiently high temperatures, both the real and the imaginary part are found to increase linearly with temperature. At optimized parameters, the temperature at which the linear behavior sets in is different for $\varepsilon_{0}(T)$ and $\gamma_{0}(T)$, being higher for the former and lower for the latter. When computed using parameters fitted from the lattice, said temperatures align and become essentially equal -- or very close to -- the deconfinement temperature: the only exception lies in the imaginary part of the pole computed in full QCD which, as reported above, still experiences a small decrease when going from $T=154$~MeV to $T=174$~MeV. While at optimized parameters $\gamma_{0}(T)$ eventually becomes larger than $\varepsilon_{0}(T)$, the calculation carried out with lattice parameters exhibits an imaginary part always remaining smaller than -- albeit of the same order of -- the real part of the pole.

The only evidence of the deconfinement phase transition borne by the gluon poles of the one-loop screened massive expansion is the linearization of their behavior with temperature for $T>T_{c}$. As we mentioned in the introduction, this feature is fully expected from deconfined quasi-particles in the high-temperature regime. We must however stress again that these poles, being located on the principal Riemann sheet of the propagator instead of secondary ones, are \textit{not} those one would ordinarily associate to quasi-particles in the deconfined phase. Whether the two kinds of poles bear any relation to one another is unclear at this stage; we will advance some hypotheses later on in our discussion.

The spectral function of the Landau-gauge gluon propagator was computed for the first time at finite temperature within a one-loop truncation of the screened massive expansion -- and more generally, as far as we know, within any perturbative massive model of QCD -- in the present study. At $T>0$, that of pure Yang-Mills theory is not dissimilar from its vacuum counterpart reported in \cite{SIR16b,SC18}: it displays positivity violation at low and high frequencies, a positivity window at intermediate frequencies, the same structure of smooth minima and maxima~-- namely, a low-frequency negative minimum, followed by positive maximum at intermediate frequencies and by two negative minima at higher frequencies --, and non-analyticities at $\omega=m,2m$, where $m$ is the gluon mass parameter. The non-analyticity at $2m$ presents itself as a discontinuity in the first derivative of the spectral function, whereas the discontinuity at $m$ only affects higher-order derivatives. As temperature increases, the spectral function is found to become overall larger in absolute value up to a temperature which, when measured using lattice parameters, is compatible with the deconfinement temperature $T_{c}\approx270$~MeV. The temperature resolution of our parameter set around $T_{c}$ is insufficient to draw conclusions on whether the two of them are the same within indeterminacies. At higher temperatures, the spectral function progressively flattens to zero.

The structure of the finite-temperature gluon spectral function computed in full QCD is much richer than in pure Yang-Mills theory due to the presence of quark mass thresholds at $\omega=2M_{q}$ in addition to $\omega=m,2m$, which generate non-analyticities for each distinct value of $M_{q}$. At low temperatures, setting $M_{q}\approx400$~MeV -- which implies $m<2M_{q}<2m$ for both the optimized parameters and for the parameters obtained in the range $T\in[139,278]$~MeV from lattice data --, the spectral function develops a two-maxima structure in the positivity region, around the lightest quark mass threshold, which has no counterpart in pure Yang-Mills theory. The maximum to the right of the threshold is dominant for very low temperatures, but it soon falls below the one located to the left at $\omega\approx m$ as temperature increases. Over the rest of its domain, the behavior of the full-QCD spectral function is not dissimilar from the pure glue case. For instance, the full-QCD spectral function too possesses a low-frequency minimum and an intermediate-frequency positive maximum~-- the latter corresponding to the aforementioned leftmost maximum in the two-maxima structure -- which, when computed using lattice parameters, are largest in absolute value at a temperature close to $T_{c}\approx 150$~MeV. This is analogous to what happens in pure Yang-Mills theory, where, as we just said, the spectral function seems to be largest around the deconfinement temperature. In full QCD, however, we are able to affirm with more confidence that the temperature where this happens is somewhat higher than $T_{c}$, and closer to $T\approx 174$~MeV.

One interesting feature of the spectral function is that~-- regardless of whether computed in pure Yang-Mills theory or full QCD -- the positions of the low-frequency minimum and, separately, those of the positive maximum, have a tendency to align across the considered temperature range as soon as one switches from temperature-independent optimized parameters to temperature-dependent parameters obtained from the lattice data. As noted in Secs.~III and IV, this is the result of temperature pushing their position to lower frequencies, compensated by the gluon mass parameter $m^{2}(T)$ pushing them to higher ones by means of its increase with $T$. If a compelling reason was found for the temperature independence of these extrema, then fixing their position could be a sensible way to determine the value of the gluon mass parameter $m^{2}(T)$ as a function of temperature without relying on the lattice.

In order to account for model-dependent effects due to keeping the quark masses fixed at all temperatures and disregarding chiral symmetry restoration and ordinary thermal mass effects, we tuned the quark masses across values ranging from $M_{q}=2$~MeV to $M_{q}=2$~GeV using optimized parameters, and found no evidence of additional poles emerging in the full-QCD gluon propagator for $T\in[0,700]$~MeV. Results obtained by lowering the quark masses to around half their value ($2M_{q}<m$) -- similar to what we did in \cite{CS25} -- suggest that, in the vicinity of $T_{c}$, 1. the real part of the full-QCD gluon pole may be enhanced in response to the decrease of the quark mass brought by chiral symmetry restoration; 2. the full-QCD spectral function as well may be enhanced in a frequency interval encompassing $\omega\in[2M_{q},m]$; 3. the shift of the quark mass threshold to lower frequencies may cause the lowest-frequency smooth minimum of the full-QCD spectral function to disappear and be replaced by a point of non-analyticity. Which of these changes actually affects the gluon propagator, and to what extent, depends on the precise evolution of the quark masses with temperature -- currently unknown in the context of the screened massive expansion.\\

Going back to our main motivation for carrying out the analysis presented in this paper, it should be apparent by now that we found no evidence of deconfinement in the form of a change of the analytic structure of the gluon propagator in the deep infrared, as computed to one loop in the Landau gauge within the screened massive expansion of QCD. In particular, we found no evidence of the formation of a sharp peak in the spectral function\footnote{This includes full QCD with $M_{q}$ in the interval $[2\text{ MeV},2\text{ GeV}]$.} at temperatures higher than $T_{c}$, mirroring a physical quasi-particle pole approaching the Minkowski axis from secondary Riemann sheets. The closest our results get to showing traces of the deconfinement transition is by displaying~-- as we already reported~-- a linear behavior with temperature for the real and the imaginary part of the poles located on the \textit{physical} Riemann sheet of the propagator for $T>T_{c}$, and minima and maxima being largest in absolute value in the spectral function at temperatures $T$ close to $T_{c}$ (with the applicable caveats for the latter, see above). These features are already present when the calculation is carried out at temperature-independent optimized parameters, but their relation with the critical temperature $T_{c}$ is only clarified after switching to temperature-dependent parameters obtained by fitting the lattice data, similar to what we observed in the static sector when studying the behavior of the chromoelectric propagator \cite{SC21,CS25}. It is important to recognize that neither of these features can be understood as either necessary or sufficient conditions for deconfinement. What they do signal is that the propagator behaves differently for $T>T_{c}$ than for $T<T_{c}$ -- that is, that something akin to a transition occurs at $T=T_{c}$ --, but they don't tell us anything about the deconfining nature itself of the transition.

The lack of clear evidence of deconfinement in our results is subject to different interpretations, the most trivial of which -- namely, the inadequacy of the one-loop screened massive expansion to provide information on the phenomena under exam -- we will leave for last. The first interpretation stems from taking our results at face value. If we assume that our results provide an accurate picture of gluodynamics, from which physical information can be directly deduced, then we must conclude that, right above the deconfinement temperature $T_{c}$, gluons still behave as confined degrees of freedom. In favor of this reading is the absence of sharp peaks in the gluon spectral function and the fact that the gluon poles are found on the principal Riemann sheet of the propagator even for $T>T_{c}$.

A second interpretation comes from the observation that the gluon propagator is a gauge-dependent object, which in covariant gauges inhabits a mathematical structure -- that of an indefinite-metric Hilbert space~-- known to contain unphysical states which do not decouple from the physical ones if not in suitable gauge-invariant matrix elements (or sums thereof). Since the gluon propagator is not one of these, it may well be that information on the physical spectrum of the gauge sector cannot be garnered by exploring its analytic structure in isolation, but require instead a combined analysis with higher-point Green functions. The analytic structure described in our study may thus reflect the exact behavior of the propagator, but its physical meaning could only be understood by computing other Green functions and combining them with the gluon propagator into suitable observables. While perfectly valid, this argument does not address the fact that the singularities of the gluon propagator are gauge invariant over wide classes of gauges, and can thus be expected -- but are not assured -- to have a physical meaning on their own.

A third possible interpretation is that the approximation provided by the screened massive expansion, especially to one loop, is insufficient to capture the thermal behavior of the gluon propagator in the deep infrared, at least in the context of the analysis carried out in the present paper. Evidence that the approach needs to be improved in the chromoelectric static sector was already discussed in Sec.~II. Here we will further note that, at high enough temperatures, non-perturbative contributions such as those borne by hard-thermal loops are known to become non-negligible. While the screened massive expansion -- especially when a temperature-dependent gluon mass parameter $m^{2}(T)$ is used for calculations -- can partially mitigate the absence of HTL-resummed contributions to the zero-order gluon propagator, the present study completely ignores other contributions, such as the HTL corrections to the 3-gluon and 4-gluon bare vertices \cite{BP90c}.

Determining which of the three interpretations above more closely approximates reality is not an easy task as it requires further investigations on a number of fronts. First of all, it necessitates a better understanding of the nature of complex-conjugate poles in the principal Riemann sheet of gauge propagators. Learning whether they are artifacts of low-order perturbation theory or actual features of gauge theories is instrumental to how we think about confinement as it relates to the information carried by the gluon propagator, and could allow us to discard results known a priori to be wrong. Second, it requires a theoretical knowledge of how physical singularities enter gauge-dependent Green functions or combinations thereof and of how they can be separated from unphysical contributions; specularly, it requires knowing how gauge-invariant singularities of such Green functions affect physical observables -- assuming that they do. Furthermore, it needs a comprehensive analysis of thermal contributions which are technically contained in higher perturbative orders, but become as important as lower orders in particular limits, such as when computing poles. This analysis was initiated in the '80s and lead to the development of the HTL formalism, but, to our knowledge, it was never carried out for QCD under the assumption that gluons also acquire a mass at zero temperature. Whatever the outcome of these investigations may be, we hope our results will have shed some light on the deep-infrared analytic structure of the gluon propagator, at least as far as a plain one-loop calculation is able to describe it.

\acknowledgments

The author is in debt with E. Salgado, O. Oliveira and P. Silva for sharing the lattice data of Ref.~\cite{SOS25}. He would like to express gratitude to F. Siringo and D. Dudal for extensive discussions on the topics of this paper, and to Dr. A. Asta, Mrs. A. Asta and Dr. A. Borys for the time spent at UDA. The author thanks the ECT* for its warm hospitality in Trento during the workshop ``The complex structure of strong interactions in Euclidean and Minkowski space'' (May 26-30, 2025), which prompted the present study. This research was supported by PIACERI ``Linea di intervento 1'' (M@uRHIC) of the University of Catania and by PRIN2022 (Projects No. 2022SM5YAS and No. P2022Z4P4B) within Next Generation EU fundings. Numerical integrations were performed using the Rust library Peroxide \cite{peroxide}.

\section*{Data availability}

The numerical results of the present study were obtained using the \texttt{qcd-sme} Rust library, and are openly available \cite{qcd-sme-rs}.

\newpage

\bibliography{ThermalQCD}

\end{document}